\documentclass[12pt]{article}
\usepackage{verbatim}
\usepackage{amsfonts}
\usepackage{graphics}
\usepackage{amsmath}
\usepackage{times}
\usepackage{appendix}
\usepackage{color}
\usepackage{enumerate}
\usepackage{fancyhdr,latexsym,amsmath,amsfonts,amssymb,amsbsy,amsthm,url}
\usepackage[margin=0.5in,footskip=0.25in]{geometry}
\usepackage{graphics,graphicx,epsfig}
\usepackage{breqn}
\usepackage{caption,subcaption,float,subfloat}

\usepackage[english]{babel}
\usepackage[dvipsnames]{xcolor}
\usepackage{tikz}
\usepackage{amsmath,amssymb,multirow,multicol}
\usepackage{caption,subcaption,float,subfloat}
\usepackage[english]{babel}
\usepackage{url}
\usepackage{hyperref}

\numberwithin{equation}{section}

\fancypagestyle{usual}{
  \lhead[\thepage]{}
  \chead[{\it \shortauthors}]{\sc \shorttitle}
  \rhead[]{\thepage}
  \lfoot[]{}
  \cfoot[]{}
  \rfoot[]{}
  
}

\makeatletter

\renewcommand{\section}{
  \@startsection
  {section}
  {1}
  {0pt}
  {1.1\baselineskip}
  {0.2\baselineskip}
  {\sc \centering}
}

\renewcommand{\subsection}{
  \@startsection
  {subsection}
  {1}
  {0pt}
  {1.1\baselineskip}
  {0.2\baselineskip}
  {\sc \centering}
}

\renewcommand{\subsubsection}{
  \@startsection
  {subsubsection}
  {1}
  {0pt}
  {1.1\baselineskip}
  {0.2\baselineskip}
  {\sc \centering}
}

\makeatother

\begin{document}

\title{\large\sc Is being `Robust' beneficial?: A perspective from the Indian market}
\normalsize
\author{\sc{Mohammed Bilal Girach} \thanks{Indian Institute of Technology Guwahati, Guwahati-781039, Assam, India, e-mail: m.girach@iitg.ac.in}
\and \sc{Shashank Oberoi} \thanks{Indian Institute of Technology Guwahati, Guwahati-781039, Assam, India, e-mail: s.oberoi@iitg.ac.in}
\and \sc{Siddhartha P. Chakrabarty} \thanks{Indian Institute of Technology Guwahati, Guwahati-781039, Assam, India, e-mail: pratim@iitg.ac.in,
Phone: +91-361-2582606, Fax: +91-361-2582649}}
\date{}
\maketitle
\begin{abstract}

The problem of data uncertainty has motivated the incorporation of robust optimization in various arenas, beyond the Markowitz portfolio optimization. This work presents the extension of the robust optimization framework for the minimization of downside risk measures, such as Value-at-Risk (VaR) and Conditional Value-at-Risk (CVaR). We perform an empirical study of VaR and CVaR frameworks, with respect to their robust counterparts, namely, Worst-Case VaR and Worst-Case CVaR, using the market data as well as the simulated data. After discussing the practical usefulness of the robust optimization approaches from various standpoints, we infer various takeaways. The robust models in the case of VaR and CVaR minimization exhibit superior performance with respect to their base versions in the cases involving higher number of stocks and simulated setup respectively.


\end{abstract}

\section{Introduction}
\label{Introduction}

The area of ``robust portfolio optimization'' has been developed primarily with the motivation of addressing the drawbacks observed for the classical mean-variance model introduced by Markowitz \cite{Markowitz1,Markowitz2}. Theoretically, the Markowitz based portfolio optimization can result in assignment of extreme weights, such as large short positions in securities comprising the portfolio. Such kind of counter-intuitive investments can be avoided by introducing appropriate constraints on the weights. However, Black and Litterman \cite{Black} argued that there is an added disadvantage in doing so, since there is a high chance of the optimal portfolio lying in the neighborhood of the imposed constraints, thus, leading to strong dependence of the constructed portfolio on the constraints. Secondly, the optimal portfolio constructed using the mean-variance analysis is highly sensitive to the estimation errors of input parameters, that include means, variances and covariances of asset returns. Based on the assumption of normal distribution of returns, the maximum likelihood estimates (sample mean and sample covariance matrix) from the historical data are used for estimating the return and risk parameters. Usually, the historical data neglects various market factors and is not an accurate representation for estimating the future returns. Taking into account the above reasons, Michaud \cite{Michaud} labeled the Markowitz model as ``estimation-error maximizers'' and argued that the mean-variance analysis tends to maximize the impact of estimation errors associated with the return and the risk parameters for the securities. As a result, the Markowitz portfolio optimization often overweighs (underweighs) the securities having higher (lower) expected return, lower (higher) variance of returns and negative (positive) correlation between their returns. Chopra and Ziemba \cite{Chopra} performed the sensitivity analysis of performance of optimal portfolios by studying the relative effect of estimation errors in means, variances and covariances of security returns, taking the investors' risk tolerance into consideration. In particular, they provided empirical evidence of high sensitivity of the Markowitz portfolio towards the expected returns of individual securities. They observed that at a high risk tolerance of around fifty, cash equivalent loss for estimation errors in means is about eleven times greater than that for errors in variances or covariances. Broadie \cite{Broadie} conducted a simulation based study to investigate the error maximization property of mean-variance analysis. He supported his argument of overestimation of expected returns of optimal portfolios obtained using the mean-variance model, through his simulated results, wherein, the estimated Markowitz efficient frontier lies above the actual frontier.

Major formulations proposed in the field of robust portfolio optimization have focused on optimizing the worst case performance of the Markowitz portfolio. These methods involve the use of uncertainty set, that includes the worst possible realizations of the uncertain input parameters, as described by Tutuncu and Koenig \cite{Tutuncu}. They have performed empirical analysis of robust allocation methods using the market data. In accordance with the observed results, they suggested robust portfolio optimization as a viable alternative for conservative investors. Significant efforts have been made towards formulating robust approaches from Markowitz based mean-variance analysis using box, ellipsoidal \cite{Fabozzi,Kim} and separable uncertainty sets \cite{Lu,Tutuncu}.

However, both the mean-variance model, as well as its robust alternatives involve the use of standard deviation, which has invited criticism. From a practitioner's point of view, the upside and downside risk cannot be perceived in the same way, because most of the times, the upside risk can improve the overall performance of the portfolio, whereas, the downside fluctuation usually results in losses that are more impactful as compared to the extent of the gains (through upside risk). Variance is not a reliable or appropriate  risk measure, if the underlying distribution is leptokurtic, which is quite often the case with real market data, where the return distribution exhibits fat tails. In order to address these issues, models involving other measures of risk, like Value-at-Risk (VaR) and Conditional Value-at-Risk (CVaR), have been developed and accordingly their corresponding robust models have been studied as well.

Unlike the mean-variance setup, VaR framework takes into account the probability of losses \cite{Var96}. Ghaoui et al. \cite{Ghaoui03} defined VaR, at confidence level $1-\epsilon$, as the minimum value of $\gamma$ such that the probability of the loss exceeding $\gamma$ is less than $\epsilon$. Though VaR takes probability of losses into account, it has its own limitations. The computational part, not only requires the knowledge of the whole distribution, but it also involves high dimensional numerical integration which may not be tractable at times. Additionally the lack of significant research on methods using Monte Carlo simulations \cite{Var96} for the design of the portfolio, is another drawback, when it comes to the applicability of VaR. Black and Litterman \cite{Black}, and Ju and Pearson \cite{ju98} discussed the issues regarding the computational difference between the true VaR and the calculated VaR, and determined that the error in the computation of the classical VaR (henceforth referred to as base VaR) can be attributed to errors in the estimation of the first and second moments of the asset returns. With the emergence of the robust formulations and optimizations, the concept of Worst-Case VaR (WVaR) not only allows for approaching the solution in a more tractable way, but also relaxes the assumptions on the information known to us apriori.  Further, we assume that only partial information about the underlying distribution is known, and consequently, the premise is that the distribution of the asset returns belongs to a family of allowable probability distributions $\mathcal{P}$ \cite{Ghaoui03}. For example, given component wise bounds of ($\hat{\boldsymbol{\mu}}$,  $\Sigma$), $\mathcal{P}$ could comprise of normally distributed random variables with $\hat{\boldsymbol{\mu}}$ and $\Sigma$ as the moments' pairs.

Despite its popularity as a measure of downside risk, it has been observed that VaR also suffers from several shortcomings \cite{Capinski_risk,Lim,Zhu}. VaR for a diversified portfolio may exceed that for an investment in a single asset. Consequently, one of the major limitations of VaR is the lack of sub-additivity in the case of general distributions. In accordance with this observation, VaR is not a coherent measure of risk as required by the definition laid out by Artzner et al. \cite{Artzner}. Secondly, VaR does not provide any information on the size of losses in adverse scenarios \textit{i.e.,} those beyond the confidence level $1-\epsilon$. VaR gives the worst of the best $1-\epsilon$ percentage of the return distribution.
Additionally, VaR is non-convex and non-differentiable when incorporated into portfolio optimization problem. As a result, it becomes difficult to optimize VaR since global minimum may not exist. CVaR, introduced by Rockafellar and Uryasev \cite{Rockafellar1,Rockafellar2}, has addressed various concerns centered around VaR. For continuous distributions, CVaR is the expected loss conditioned on the loss outcomes exceeding VaR. It gives the average of the worst $\epsilon$ percentage of returns. It has been proved by Pflug \cite{Pflug}, and Acerbi and Tasche \cite{Acerbi} that CVaR is a coherent measure of risk and hence it satisfies the property of sub-additivity. Therefore, CVaR can be minimized through investment in a diversified portfolio. Unlike VaR, CVaR takes into consideration the impact of losses beyond the confidence level $1-\epsilon$ \cite{Capinski_risk}. Also, the minimization of CVaR is a convex optimization problem \cite{Lim}. Similar to Markowitz optimization and VaR minimization, there is an issue of lack of robustness in the classical framework of CVaR minimization. Computing CVaR requires the complete knowledge of the return distribution as argued by Zhu and Fukushima \cite{Zhu}. They have addressed this shortcoming of the classical CVaR (henceforth, referred to as base CVaR) by proposing a new risk measure, namely, Worst-Case CVaR (WCVaR).

Motivated by the progress made towards incorporating robust optimization in the framework of risk minimization, this paper focuses on assessing the practical usefulness of the robust counterparts for the classical formulations of VaR and CVaR minimization. Accordingly, we perform empirical analysis of the performance of the base VaR and base CVaR models with respect to their robust formulations, WVaR and WCVaR models using both market and simulated data. Additionally, we provide relevant insights regarding the viability of these robust models over their classical formulations from the perspective of an investment practitioner. For performing empirical study, we have used the data obtained from the Indian indices of
S\&P BSE 30 and S\&P BSE 100.

The rest of the paper is organized as follows. In Section \ref{Risk_Minimization_Approaches}, we discuss the tractable formulations of the downside risk minimization methods used in the work. Section \ref{Computational_Results} presents the empirical results observed on comparison of the performance of robust risk minimization models with their classical counterparts. Section \ref{Discussion} focuses on analyzing the practical viability of the robust models from the standpoint of number of stocks, sample size and type of data. In Section \ref{Conclusion}, we sum up the key inferences from this work.

\section{Risk Minimization Approaches}
\label{Risk_Minimization_Approaches}

In this section, we discuss the mathematical formulations for the problems of optimizing VaR and CVaR along with their robust counterparts, namely,
WVaR and WCVaR respectively. Accordingly, we begin with a glossary of notation (to be used in the mathematical formulation) as follows:
\begin{enumerate}
\item $N$: The total number of assets in the portfolio.
\item $\displaystyle{\mathbf{x}}$: Vector of weights of assets comprising the portfolio.
\item $\displaystyle{\boldsymbol{\mu}}$: Vector for expected returns of assets.
\item $\displaystyle{\Sigma}$: Covariance matrix for returns of assets.
\item $\displaystyle{\boldsymbol{r}}$: Random return vector for assets.
\item $\displaystyle{\boldsymbol{{\hat{\mu}}}}$: Estimate for $\displaystyle{\boldsymbol{\mu}}$.
\end{enumerate}

\subsection{Value-at-Risk (VaR)}

Recall that the VaR, at confidence level $1-\epsilon$, is the minimum value of $\gamma$ such that the probability of loss exceeding $\gamma$ is less
than $\epsilon$. Accordingly, the VaR is mathematically defined as,
\begin{equation}
VaR_{\epsilon}(\mathbf{x})=\min\left\{\gamma:P\left\{\gamma \leq -r(\mathbf{x},\boldsymbol{\mu})\right\} \leq \epsilon \right\},
\label{fig:var_basic}
\end{equation}
where $\epsilon \in (0,1)$. When we deal with mean-variance setup, only the mean and the variance \textit{i.e.,} the first and the second moments of the asset returns are required, but for the computation of VaR, it is necessary to have the knowledge of the entire distribution. If the underlying distribution is Gaussian with the moments' pair as $\displaystyle\left(\hat{\boldsymbol{\mu}},\Sigma\right)$, then VaR can be computed via the following analytical form:
\begin{equation}
VaR_{\epsilon}(\mathbf{x})=\kappa(\epsilon)\sqrt{\mathbf{x}^{\top}\Sigma \mathbf{x}}-\hat{\boldsymbol{\mu}}^{\top}\mathbf{x},
\label{eqn:kappa_eqn}
\end{equation}
where $\displaystyle{\kappa(\epsilon)=-\Phi^{-1}(\epsilon)}$ with $\Phi(z)$ representing the cumulative distribution for standard normal random variable.
In practice, the value of the $\displaystyle{\kappa(\epsilon)}$ can be determined only if the cumulative distribution function of the underlying distribution is known apriori. However, if the distribution is unknown, then we have to rely on the Chebyshev's inequality. The bound (upper) due to Chebyshev's inequality, only requires the knowledge of the first two moments' pair. We call a bound to be exact if the upper bound is attained. If not, we use the following bound, given by Bertsimas and Popescu \cite{Bert05},
\begin{equation}
\kappa(\epsilon)=\sqrt{\frac{1-\epsilon}{\epsilon}}.
\label{fig:kappa_epsilon_def}
\end{equation}
Accordingly, we formulate the problem of minimization of the generalized VaR as follows:
\begin{equation}
\min \kappa(\epsilon)\sqrt{\mathbf{x}^{\top}\Sigma \mathbf{x}}-\hat{\boldsymbol{\mu}}^{\top}\mathbf{x},~\text{subject to}~
\mathbf{x} \in \mathcal{X},
\label{fig:var_general}
\end{equation}
where $\displaystyle{\kappa(\epsilon)}$ is an appropriate factor of risk chosen in accordance with the underlying distribution of asset returns and
$\displaystyle{\mathcal{X}=\left\{\mathbf{x}: \mathbf{x}^{\top}\mathbf{1}=1~\text{and}~\mathbf{x} \geq 0 \right\}}$.
The function $VaR_{\epsilon}(\mathbf{x})$ is convex and the global optimum can be obtained using techniques like interior-point methods and
second order cone programming (SOCP).

\subsection{Worst-Case VaR (WVaR)}

Given a probability level $\epsilon$, the worst-case VaR can be formulated as,
\begin{equation}
V_{\epsilon}^{\mathcal{P}}(\mathbf{x})=\min\left\{\gamma: \sup_{P \in \mathcal{P}} P\left\{\gamma \leq -r(\mathbf{x},\boldsymbol{\mu})\right\}
\leq \epsilon \right\},
\label{fig:wc_var_basic}
\end{equation}
where $\mathcal{P}$ is the family of allowable probability distributions. Accordingly, the robust formulation for WVaR minimization can be written as,
\begin{equation}
WVaR_{\epsilon}(\mathbf{x^{*}})=\min V_{\epsilon}^{\mathcal{P}}(\mathbf{x}),~\text{subject to}~\mathbf{x} \in \mathcal{X}.
\label{fig:wc_var_general}
\end{equation}
The above optimization problems can be computed by a semi-definite programming (SDP) problem which again uses the interior-point methods, as mentioned in the case of VaR. We deal with the above finite dimensional problems by using bundle methods which are mainly used for large-scale (sparse) problems. Motivated by the work on separable uncertainty sets \cite{Tutuncu}, one can view the robust formulation given by Ghaoui et al. \cite{Ghaoui03} in case of Polytopic uncertainty, as a robust formulation of WVaR involving separable uncertainty.
Accordingly, the formulation of the optimization problem for WVaR is:
\begin{equation}
\min \kappa(\epsilon)\sqrt{\mathbf{x}^{\top}\overline{\Sigma}\mathbf{x}}-\underline{\hat{\boldsymbol{\mu}}}^{\top}\mathbf{x},~\text{subject to}~
\mathbf{x} \in \mathcal{X},
\label{fig:var_poly}
\end{equation}
where $\overline{\Sigma}$ and $\underline{\hat{\boldsymbol{\mu}}}$ are the upper bound for covariance matrix and the lower bound for the
estimated mean of asset returns, respectively. These values are obtained by using the Non-Parametric Bootstrap algorithm where the type of distribution is unknown. This can also be viewed as a robust formulation involving polytopic uncertainty because ``Separable'' uncertainty is a special case of ``Polytopic'' uncertainty. In case of models involving ellipsoidal uncertainty sets, the robust WVaR formulation is not very trivial and such models mainly revolve around the
assumption of factor models.

\subsection {Conditional Value-at-Risk (CVaR)}

Recall that for continuous distributions, CVaR is the expected loss conditional on the loss outcomes exceeding VaR.
Therefore, for $\epsilon \in (0,1)$ , CVaR, at confidence level $1-\epsilon$, is defined as:
\begin{equation}
\label{eq:6.1}
CVaR_{\epsilon}(\mathbf{x}) \triangleq \frac{1}{\epsilon} \, \int \limits_{-\mathbf{r}^{\top}\mathbf{x} \geq VaR_{\epsilon}(\mathbf{x})} -\mathbf{r}^{\top}\mathbf{x} \, \, p(\mathbf{r})  d\mathbf{r},
\end{equation}
where $\mathbf{r}$ is the random return vector having probability density function $p(\mathbf{r})$ and $\mathbf{x}$ is the weight vector for a portfolio.
As proved by Rockafeller and Uryasev \cite{Rockafellar1} , $CVaR_{\epsilon}(\mathbf{x})$, defined in equation (\ref{eq:6.1}), can be transformed into:
\begin{equation}
CVaR_{\epsilon}(\mathbf{x})=\min_{\gamma \in \mathcal{R}^{N}} F_{\epsilon}(\mathbf{x},\gamma),
\label{eq:6.4}
\end{equation}
where $N$ is number of assets in the portfolio and $F_{\epsilon}(\mathbf{x},\gamma)$ is defined as:
\begin{equation}
F_{\epsilon}(\mathbf{x},\gamma) \triangleq \gamma+\frac{1}{\epsilon} \int \limits_{\mathbf{r} \in \mathcal{R}^{N}}
\left[-\mathbf{r}^{\top}\mathbf{x}-\gamma\right]^{+} p(\mathbf{r}) d\mathbf{r}.
\label{eq:6.5}
\end{equation}
In the above equation, $\left[t\right]^{+}=\max \left\{t,0\right\}$. The problem of approximating the integral involved in equation (\ref{eq:6.5})
can be dealt by sampling the probability distribution of $\mathbf{r}$ according to its density $p(\mathbf{r})$. Assuming there are $S$ samples,
$\displaystyle{\left\{\mathbf{r}_{1},\mathbf{r}_{2},\dots,\mathbf{r}_{i},\dots,\mathbf{r}_{S}\right\}}$, for the return vector $\mathbf{r}$,
$F_{\epsilon}(\mathbf{x},\gamma)$ can be approximated as \cite{Rockafellar1}:
\begin{equation}
F_{\epsilon}(\mathbf{x},\gamma) \approx \gamma+\frac{1}{S\epsilon}\sum\limits_{i=1}^{S} \left[-\mathbf{r}_{i}^{\top}\mathbf{x}-\gamma\right]^{+}.
\label{eq:6.6}
\end{equation}
In accordance with the above approximation, the problem of minimization of the base CVaR, assuming no short-selling constraints, can be formulated as the following Linear Programming Problem (LPP) \cite{Rockafellar1,Zhu}:
\begin{eqnarray}
&&\min_{(\mathbf{x},\mathbf{u},\gamma,\theta)}\theta~\text{such that}~\nonumber \\
&&\mathbf{x}^{\top}\mathbf{1}=1,~\mathbf{x} \geq \mathbf{0}, \nonumber \\
&&\gamma+\frac{1}{S\epsilon} \mathbf{1}^{\top}\mathbf{u} \leq \theta, \nonumber \\
&&u_{i} \geq -\mathbf{r}_{i}^{\top}\mathbf{x}-\gamma,~u_{i} \geq 0,~ i=1,2,\dots,S.
\label{eq:6.7}
\end{eqnarray}
where the auxiliary vector $\mathbf{u} \in \mathcal{R}^S$ and $\theta$ is the auxiliary variable that is optimized to obtain the optimal value of base CVaR.

\subsection {Worst-Case CVaR (WCVaR)}

For a fixed weight vector $\mathbf{x}$, at confidence level $1-\epsilon$, WCVaR is defined as,
\begin{equation}
WCVaR_{\epsilon}(\mathbf{x}) \triangleq \sup_{p(\mathbf{r}) \in \mathcal{P}}CVaR_{\epsilon}(\mathbf{x}),
\label{eq:6.2}
\end{equation}
where we assume that the density function $p(\mathbf{r})$ of returns belongs to a set $\mathcal{P}$ of probability distributions. Zhu and Fukushima \cite{Zhu} have proved the coherence of WCVaR as a risk measure by analyzing it in terms of worst-case risk measure $\rho_{w}$, given by,
\begin{equation}
\rho_{w}(X) \triangleq \sup_{p(\mathbf{r}) \in \mathcal{P}} \rho (X).
\label{eq:6.3}
\end{equation}
We skip the discussion on WCVaR formulations using the box uncertainty set and the ellipsoidal uncertainty set, since they require a set of possible return distributions to be assumed, so as to obtain bounds and scaling matrix respectively. However, our problem setup doesn't involve a set of return distributions, since we make use of only market data (where return distribution is not known) and simulated data (where we generate a known return distribution).
We describe the formulation of the optimization problem involving WCVaR, using mixture distribution uncertainty, by assuming that the return distribution belongs to a set of distributions
comprising of all possible mixtures of some prior likelihood distributions \cite{Zhu}. Mathematically, it is assumed that:
\begin{equation}
p(\mathbf{r}) \in \mathcal{P}_{M} \triangleq \left\{\sum\limits_{j=1}^{l} \lambda_{j}p^{j}(\mathbf{r}):\sum\limits_{j=1}^{l}\lambda_{j}=1,
\lambda_{j} \geq 0,j=1,2,\dots,l \right\}.
\label{eq:6.8}
\end{equation}
In the above equation, $p^{j}(\mathbf{r})$ denotes the $j^{th}$ likelihood distribution and $l$ is the number of the likelihood distributions.
In accordance with above assumption of mixture distribution uncertainty (that involves $\mathcal{P}_{M}$ as a compact convex set),
$WCVaR_{\epsilon}(\mathbf{x})$, defined in equation (\ref{eq:6.2}), can be rewritten as the following min-max problem \cite{Zhu}:
\begin{eqnarray}
&& WCVaR_{\epsilon}(\mathbf{x})=\min_{\alpha \in \mathcal{R}} \max_{j \in \mathcal{L}} F_{\epsilon}^{j}(\mathbf{x},\gamma),~\text{where}, \nonumber \\
&& \mathcal{L} \triangleq \left\{1,2,\dots,l\right\}, \nonumber \\
&& F_{\epsilon}^{j}(\mathbf{x},\gamma) \triangleq \gamma+\frac{1}{\epsilon}\int \limits_{\mathbf{r} \in \mathcal{R}^{N}}
\left[-\mathbf{r}^{\top}\mathbf{x}-\gamma\right]^{+} p^{j}(\mathbf{r}) d\mathbf{r}.
\label{eq:6.9}
\end{eqnarray}
Similar to the case involving the classical CVaR in the preceding Subsection, $\displaystyle{F_{\epsilon}^{j}(\mathbf{x},\gamma)}$
can be approximated via discrete sampling as,
\begin{equation}
F_{\epsilon}^{j}(\mathbf{x},\gamma) \approx \gamma+\frac{1}{S_{j}\epsilon}\sum\limits_{i=1}^{S_{j}}
\left[-\mathbf{r}_{i,j}^{\top}\mathbf{x}-\gamma\right]^{+}.
\label{eq:6.10}
\end{equation}
In the above equation, $\mathbf{r}_{i,j}$ is the $i^{th}$ sample of the return with respect to $j^{th}$ likelihood distribution and $S_{j}$ is the
number of samples corresponding to $j^{th}$ likelihood distribution. Accordingly, assuming non-negative weights, the problem of minimization of WCVaR over a feasible set of portfolios can be formulated as the following LPP \cite{Zhu},
\begin{eqnarray}
&& \min_{(\mathbf{x},\mathbf{u},\gamma,\theta)} \theta,~\text{such that} \nonumber \\
&& \mathbf{x}^{\top}\mathbf{1}=1,~\mathbf{x} \geq \mathbf{0}, \nonumber \\
&& \gamma + \frac{1}{S_{j} \epsilon} \mathbf{1}^{\top}\mathbf{u}^{j} \leq \theta, j=1,2,\dots,l, \nonumber \\
&& u_{i}^{j} \geq-\mathbf{r}_{i,j}^{\top}\mathbf{x}-\gamma,~u_{i}^{j} \geq 0,~i=1,2,\dots,S_{j},~j=1,2,\dots,l.
\label{eq:6.11}
\end{eqnarray}
In the above equation, the auxiliary vector $\mathbf{u}=(\mathbf{u}^{1};\mathbf{u}^{2};\dots; \mathbf{u}^{l}) \in \mathcal{R}^{S}$, where $\displaystyle{S=\sum\limits_{j=1}^{l}S_{j}}$, and $\theta$ is the auxiliary variable whose optimization yields the optimal value of the WCVaR.

\section{Computational Results}
\label{Computational_Results}

In this Section, we present the computational results in detail. A comparative analysis of the base VaR and the base CVaR models, with their respective worst case counterparts,
namely WVaR and WCVaR, is presented both in case of $N=31$ and $N=98$ assets.

\subsection{VaR vis-a-vis WVaR}

We carry out a comparative analysis on the performance of the optimal portfolios obtained when VaR (referred as ``base VaR model'' from hereon) and WVaR are incorporated as measures of risk in the robust portfolio optimization problem. In order to take into account only the downside risk, along with excess return for measuring the performance of a portfolio, we use the Sortino Ratio (SR) which considers the semi-deviation, denoted by $\sigma^{d}$. Since the yield for Treasury Bill in India from 2016 to 2018 was observed to be around $6\%$ \cite{Rbi}, therefore, we assumed the annualized risk-free rate to be equal to $6\%$.

We perform the empirical analysis for the available historical data for S\&P BSE 30 $(N=31)$ and S\&P BSE 100 $(N=98)$. For the first scenario $(N=31)$, we use the daily log-returns based on daily adjusted closing price of the $31$ stocks comprising S\&P BSE 30 (data source: Yahoo Finance \cite{Yf}). Accordingly, we have considered the period from December 18, 2017 to September 30, 2018 (both inclusive) comprising of a total of $194$ active trading days. For the second scenario $(N=98)$, we use the log-returns based on daily adjusted close price data of the $98$ stocks comprising S\&P BSE 100 (data source: Yahoo Finance \cite{Yf}) with the period spanning from December 18, 2016 to September 30, 2018. The reason behind this setup is to observe the trends/patterns in the performance of the portfolio when the number of stocks taken into consideration for constructing the optimal portfolio are changed. Furthermore, we also analyze the performance of the portfolio, when instead of real market data, simulated data with true moments' pair is fed to the robust optimization problem. We simulate this using Non-parametric Bootstrap Algorithm which involves sampling with ``replacement'' at $95\%$ confidence level. We delve one step further into the above setup and vary the number of simulations to observe the changes in the performance of the optimal portfolio by increasing the number of simulations. For the above case, we simulate two data-sets with true moments' pair, one with exact number of samples as in the available real market data, say, $\zeta(< 1000)$ and another with a large number of samples, say $1000$. We also use the same setup in order to analyze the performance of the portfolio when different type of data is taken into consideration \textit{i.e.,} real market and simulated data. For the computational part of the robust formulation, we choose $\mathcal{P}$ to be family of distribution with the first and the second moments as the true mean and variance of the historical data. As the knowledge of the distribution is unknown, we use $\kappa$ from equation (\ref{fig:kappa_epsilon_def}), where we only consider values of $\epsilon$ lying in $(0,0.1)$ \textit{i.e.,} the confidence level is greater than or equal to $90\%$. In the following subsections, we analyze each scenario in detail with appropriate figures and tables.

\subsubsection{Performance with $N=31$ Assets}

\begin{figure}[!h]
\centering
\includegraphics[height=7.0cm]{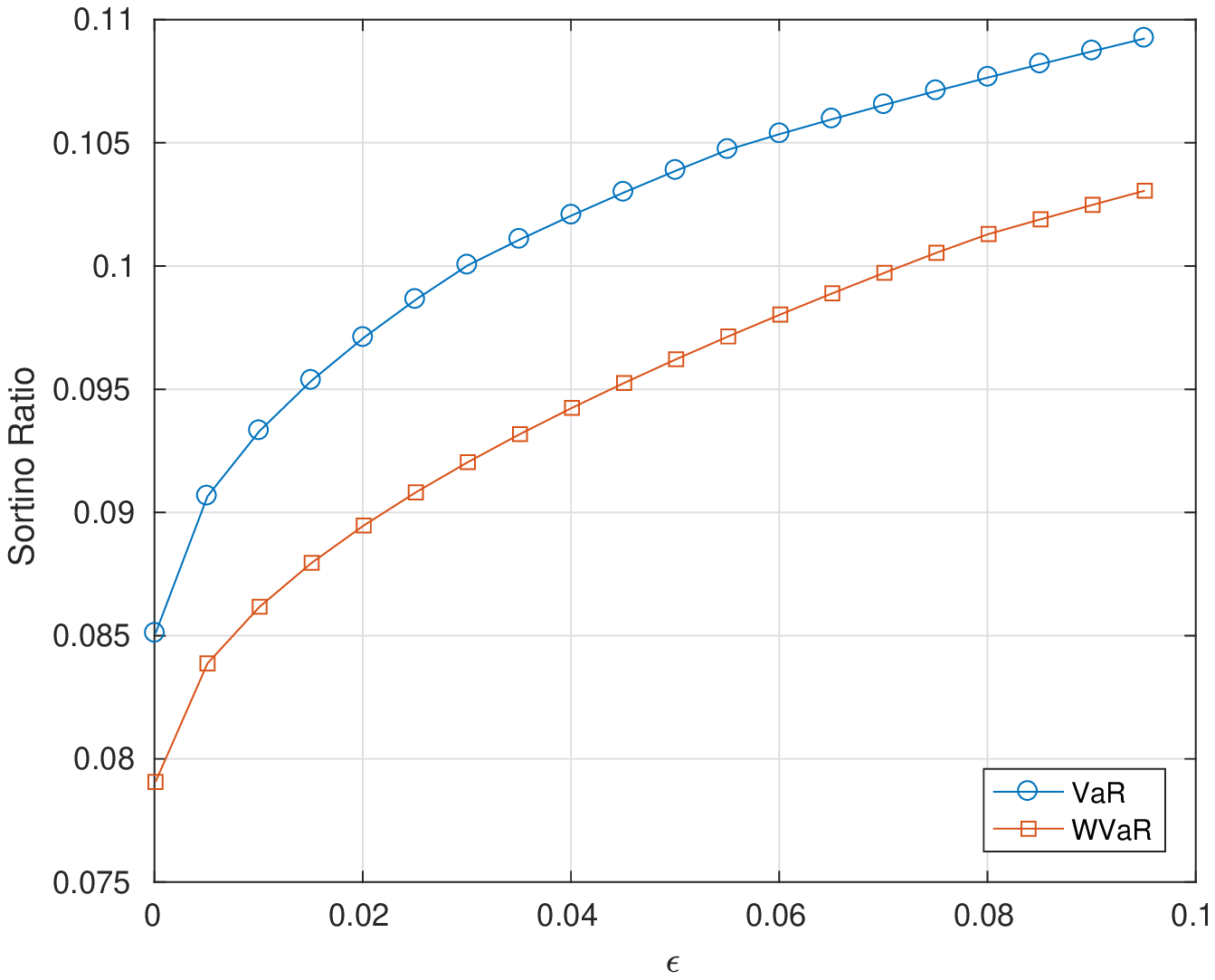}
\caption{Sortino ratio plot for base VaR and WVaR models in case of market data (31 assets).}
\label{fig:5.1}
\end{figure}

We begin by analyzing the performance of the portfolio when real market data is used to construct the optimal portfolio. A comparative analysis is carried out on the basis of the pattern exhibited by the plot of the Sortino Ratio and tabulation of results for some selected values of $\epsilon$. From Figure \ref{fig:5.1}, one can observe that the performance of the portfolio is better when the base VaR is used in the optimization problem, with the base VaR model outperforming the WVaR model for $\epsilon \in (0,0.1)$. For a better understanding of the trends, we tabulated the values of Sortino Ratio of the optimal portfolios obtained for different values of $\epsilon$, in Table \ref{tab:5.1}. We can observe that the average value of Sortino Ratio is more for the base VaR model when compared to the WVaR model, which is evident from Figure \ref{fig:5.1} as well.

\begin{figure}[!h]
\centering
\includegraphics[height=7.0cm]{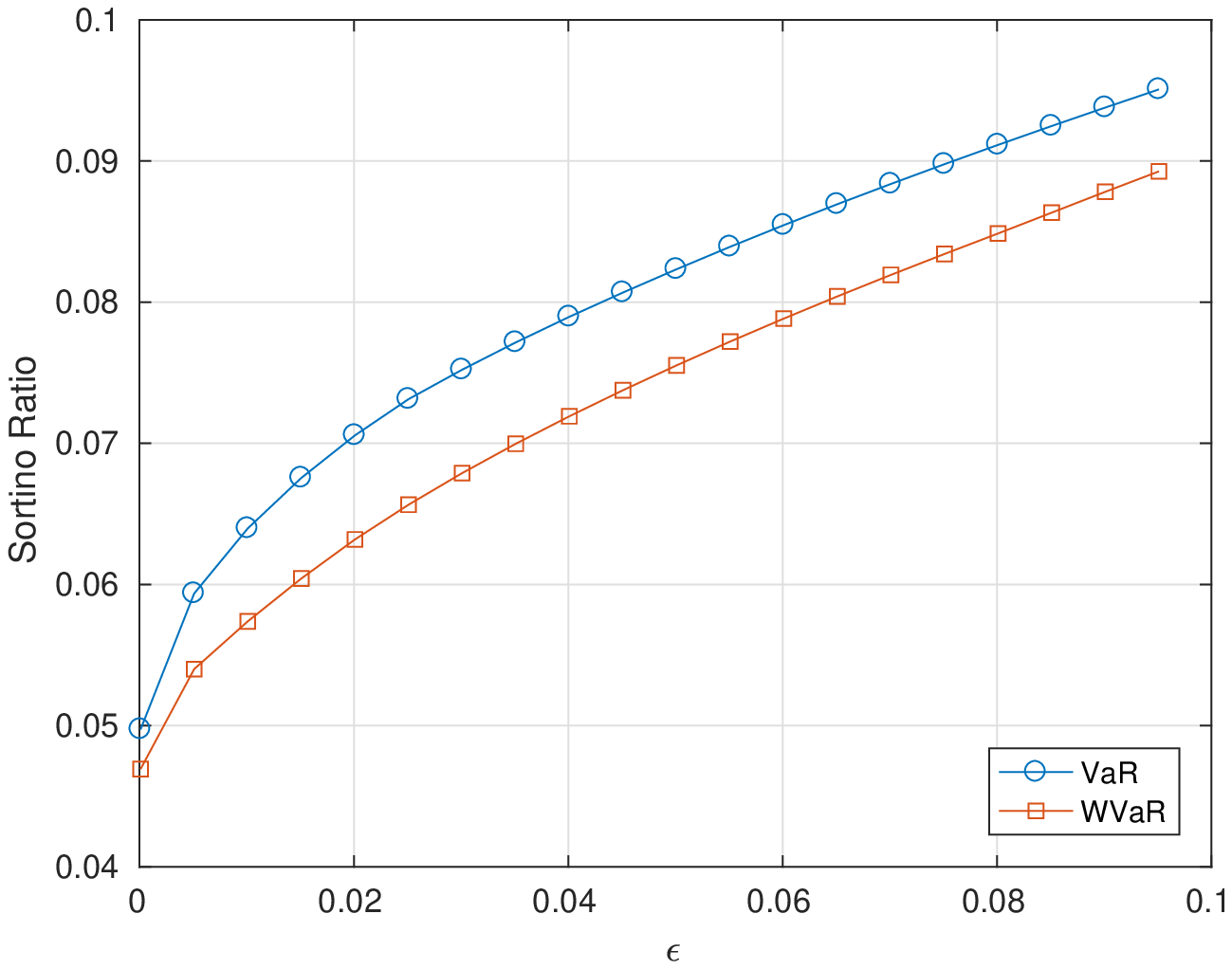}
\caption{Sortino ratio plot for base VaR and WVaR models in case of simulated data with $\zeta$ number of samples (31 assets).}
\label{fig:5.2}
\end{figure}

Now, we move on to the case of the simulated data. Firstly, we discuss the trends in the performance of the optimal portfolio when the number of the samples ($\zeta$) generated are the same as the real market data. We observe from the Figure \ref{fig:5.2} that there is little difference in performance trends, when we used $\zeta$ number of simulations, as compared to when we used the real market data. Also from Table \ref{tab:5.2}, we can observe that the base VaR model exhibits superior performance over WVaR model over the complete range of $\epsilon$.

\begin{figure}[!h]
\centering
\includegraphics[height=7.0cm]{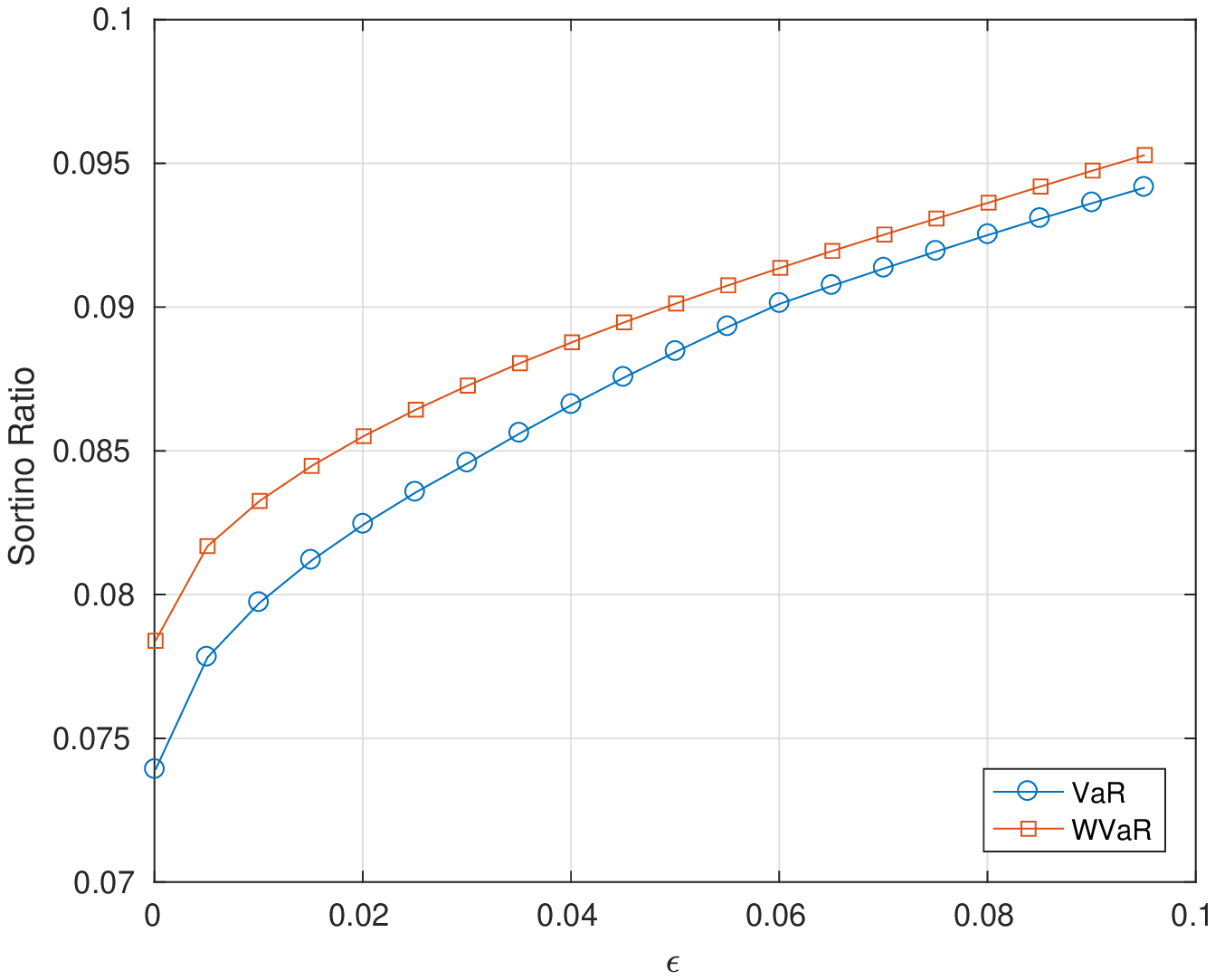}
\caption{Sortino ratio plot for base VaR and WVaR models in case of simulated data with $1000$ samples (31 assets).}
\label{fig:5.3}
\end{figure}

Since the model with $\zeta$ number of samples results in the same trends as in the real market data, we now consider the case where we simulated larger number of samples ($1000$), a choice made on the premise that larger the number of samples, less is the difference between true moments' pair and their computed counterparts. In this case, from Figure \ref{fig:5.3}, we observe that the WVaR model exhibits better performance than the base VaR model in the whole $\epsilon$ interval of $(0,0.1)$. This is suggestive that in case of simulated environment with larger number of samples, the WVaR model will be more beneficial than the corresponding VaR model. In Table \ref{tab:5.3}, one can validate the observation made above, the WVaR model performing better than the base VaR model in the entire $\epsilon$ interval of $(0,0.1)$. So, we conclude that in this case, WVaR model outperforms VaR model from the perspective of a rational investor.

We now extend a similar type of analysis for a larger set of stocks for which we obtain data from S\&P BSE 100 ($N = 98$).

\subsubsection{Performance with $N=98$ Assets}

\begin{figure}[!h]
\centering
\includegraphics[height=7.0cm]{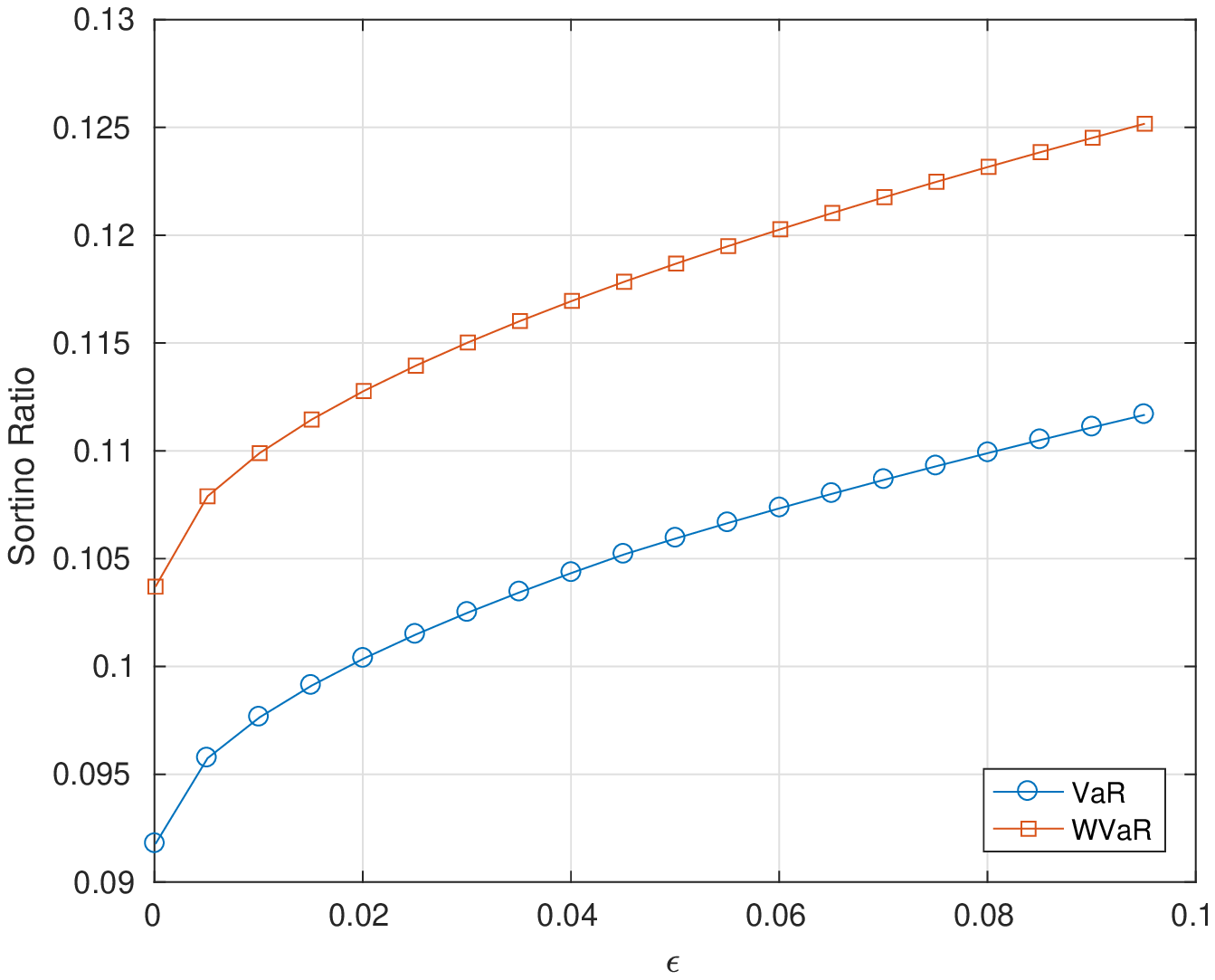}
\caption{Sortino ratio plot for base VaR and WVaR models in case of market data (98 assets).}
\label{fig:5.4}
\end{figure}

We now analyze the performance of the base VaR and WVaR models by using real market data from S\&P BSE 100. We present the empirical results in Figure \ref{fig:5.4} and tabulate the observations in Table \ref{tab:5.4}. One can observe from Figure \ref{fig:5.4} that the WVaR model exhibits superior performance in comparison to the corresponding base VaR model on the entire range of $\epsilon$. This observation can also be quantitatively justified by Table \ref{tab:5.4} where the Sortino Ratio of the portfolios obtained from WVaR model is greater than that of base VaR model.

\begin{figure}[!h]
\centering
\includegraphics[height=7.0cm]{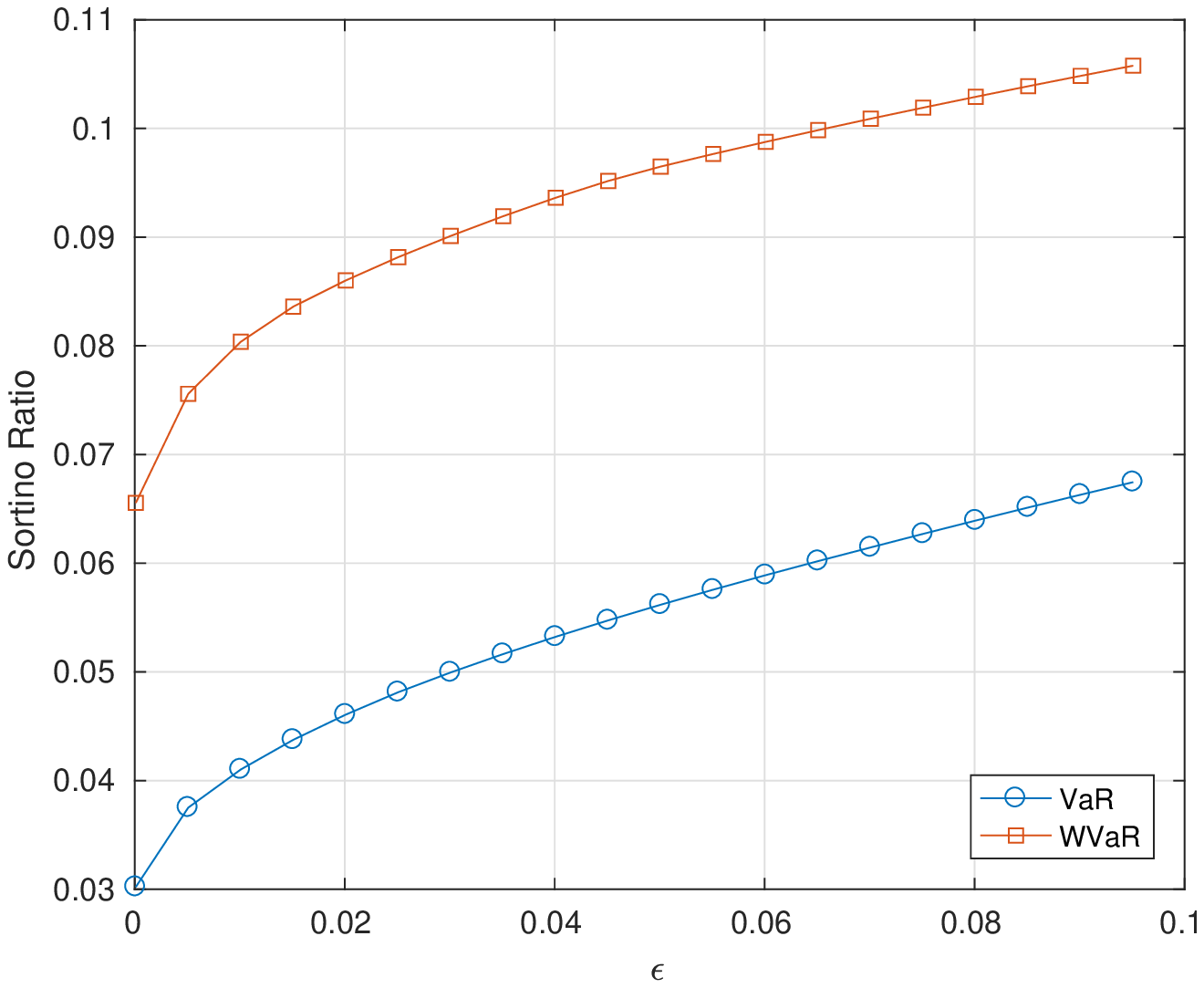}
\caption{Sortino ratio plot for base VaR and WVaR models in case of simulated data with $\zeta$ number of samples (98 assets).}
\label{fig:5.5}
\end{figure}

Similar trend is observed for $N=98$ stocks, in the case of simulated data as well, when the number of simulations equals $\zeta$. From Figure \ref{fig:5.5} and the Table \ref{tab:5.5}, we infer that the WVaR model performs better than the base VaR model, when Sortino ratio is used as performance measure, for every $\epsilon \in$ (0,0.1).

\begin{figure}[!h]
\centering
\includegraphics[height=7.0cm]{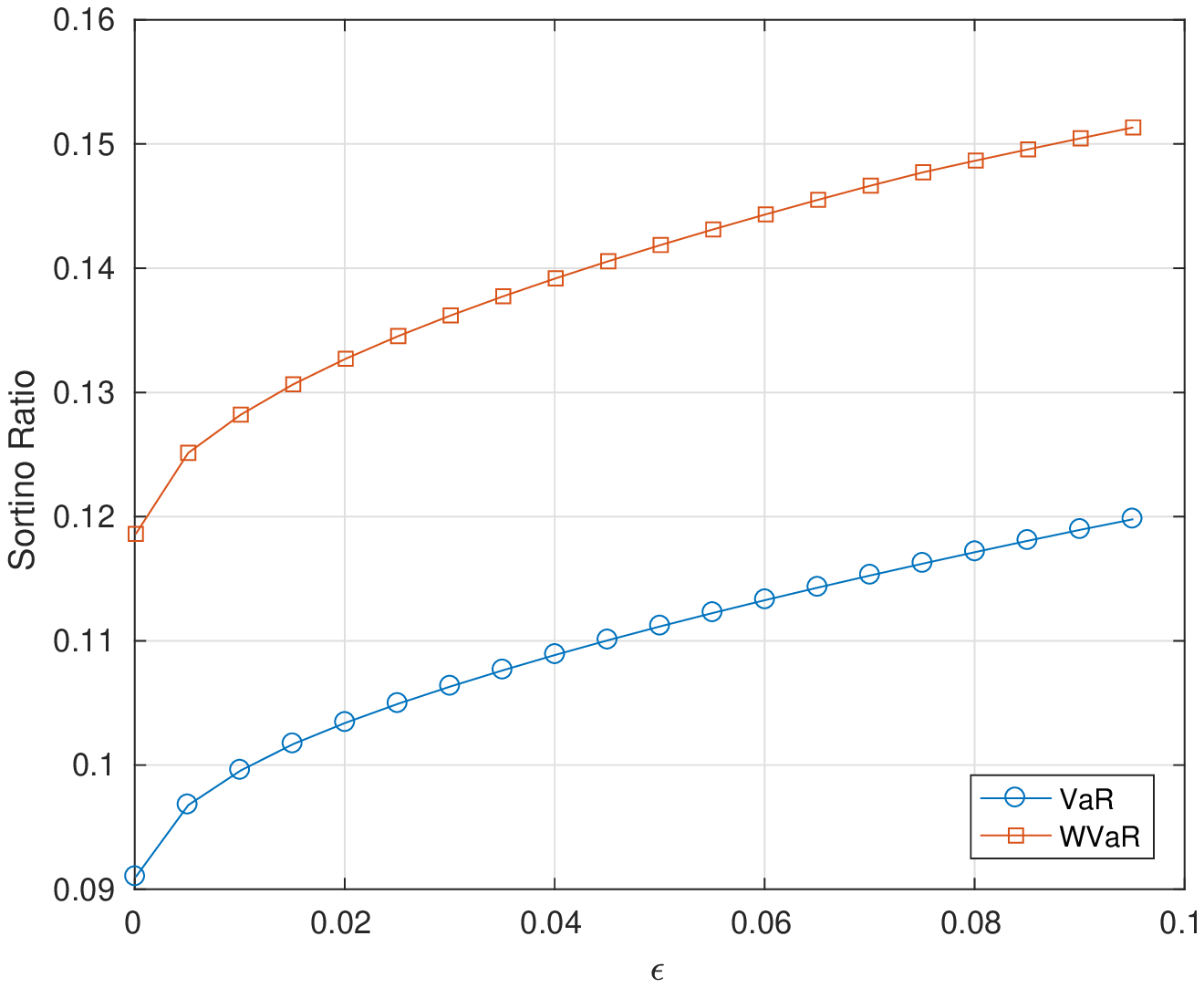}
\caption{Sortino ratio plot for base VaR and WVaR models in case of simulated data with $1000$ samples (98 assets).}
\label{fig:5.6}
\end{figure}

Finally, we consider the case where we simulated larger number of samples in order to distinguish the fluctuations in the performance of the portfolio when number of simulations are varied. Similar kind of inferences can be drawn from Figure \ref{fig:5.6} and Table \ref{tab:5.6}. The optimal portfolios obtained from the WVaR model have greater values of Sortino ratio as compared to those obtained from base VaR model, irrespective of the value of $\epsilon$.

\subsection{CVaR vis-a-vis WCVaR}

Similar to the computational analysis of robust methods in VaR minimization, we use the Sortino Ratio to perform empirical analysis of the WCVaR model vis-a-vis the base CVaR model,
using the same three sets of data.

As before, during computation for the base CVaR model, $S$ in equation (\ref{eq:6.7}) is set equal to the number of return samples, \textit{i.e,} either $\zeta$ or $1000$, depending upon the set of data used. We perform computation of the WCVaR model, as formulated in equation (\ref{eq:6.11}), for values of
$l \in \{2,3,4,5\}$ by setting $\displaystyle{S_{j}=\frac{S}{l}}$ where $\displaystyle{j \in \{1,2,\dots,l\}}$. Note that $l=1$ yields the same case as the base CVaR model. We do not include larger values of $l$ in the discussion on the empirical results since this leads to a decrease in the sample size for the likelihood distributions.

In the following subsections, we present the computational results for the two scenarios.

\subsubsection{Performance with $N=31$ Assets}

\begin{figure}[!h]
\centering
\includegraphics[height=7.0cm]{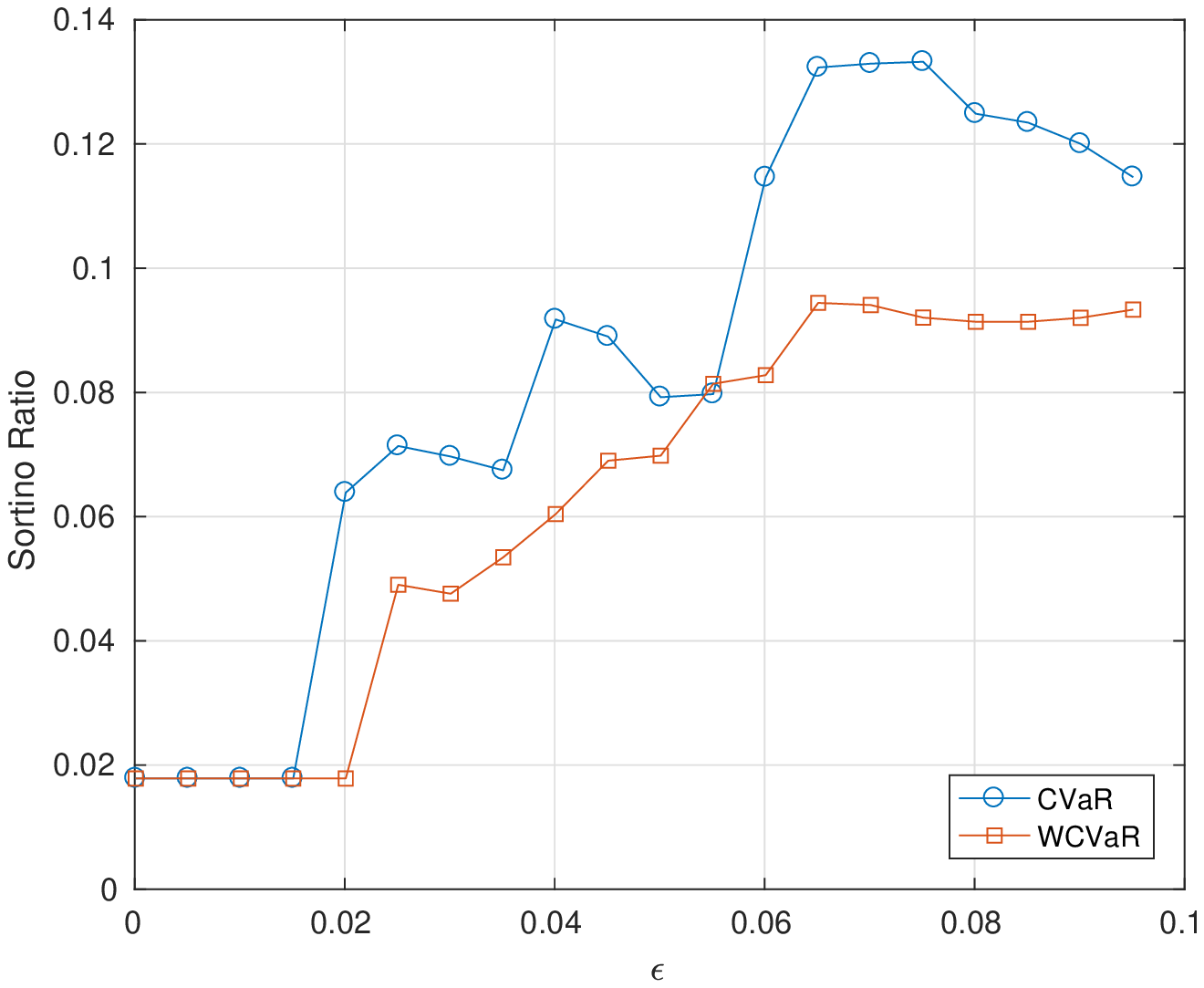}
\caption{Sortino ratio plot for base CVaR and WCVaR models in case of market data (31 assets) for $l=2$.}
\label{fig:6.1}
\end{figure}

We begin with the analysis for $N=31$ assets, in the case of the historical market data having log-returns of the stocks comprising S\&P BSE 30. In Table \ref{avgtab:6.1}, we present the average Sortino Ratio of the portfolios constructed for the base CVaR and WCVaR models (denoted by $Avg. SR_{CVaR}$ and $Avg. SR_{WCVaR}$, respectively) for different values of $l$. \textbf{In order to capture the maximum potential of WCVaR, for this case as well as other cases, we choose the value of $l$ with maximum difference in the average Sortino Ratio between the WCVaR and CVaR model.} Accordingly, we perform comparative study based on the plot of the Sortino Ratio and tabulation of results for some selected values of $\epsilon$. The motivation behind the adoption of this kind of methodology for empirical analysis is to assess the practical viability of the robust technique over the classical method in CVaR minimization. In accordance with our methodology, the empirical results are presented for $l=2$ in Figure \ref{fig:6.1} and Table \ref{tab:6.1}. From Figure \ref{fig:6.1}, we observe that the base CVaR model performs better than its robust counterpart in terms of the Sortino Ratio for the constructed portfolios with $\epsilon \in (0,0.1)$. Table \ref{tab:6.1} supports the above observation quantitatively.

\begin{figure}[!h]
\centering
\includegraphics[height=7.0cm]{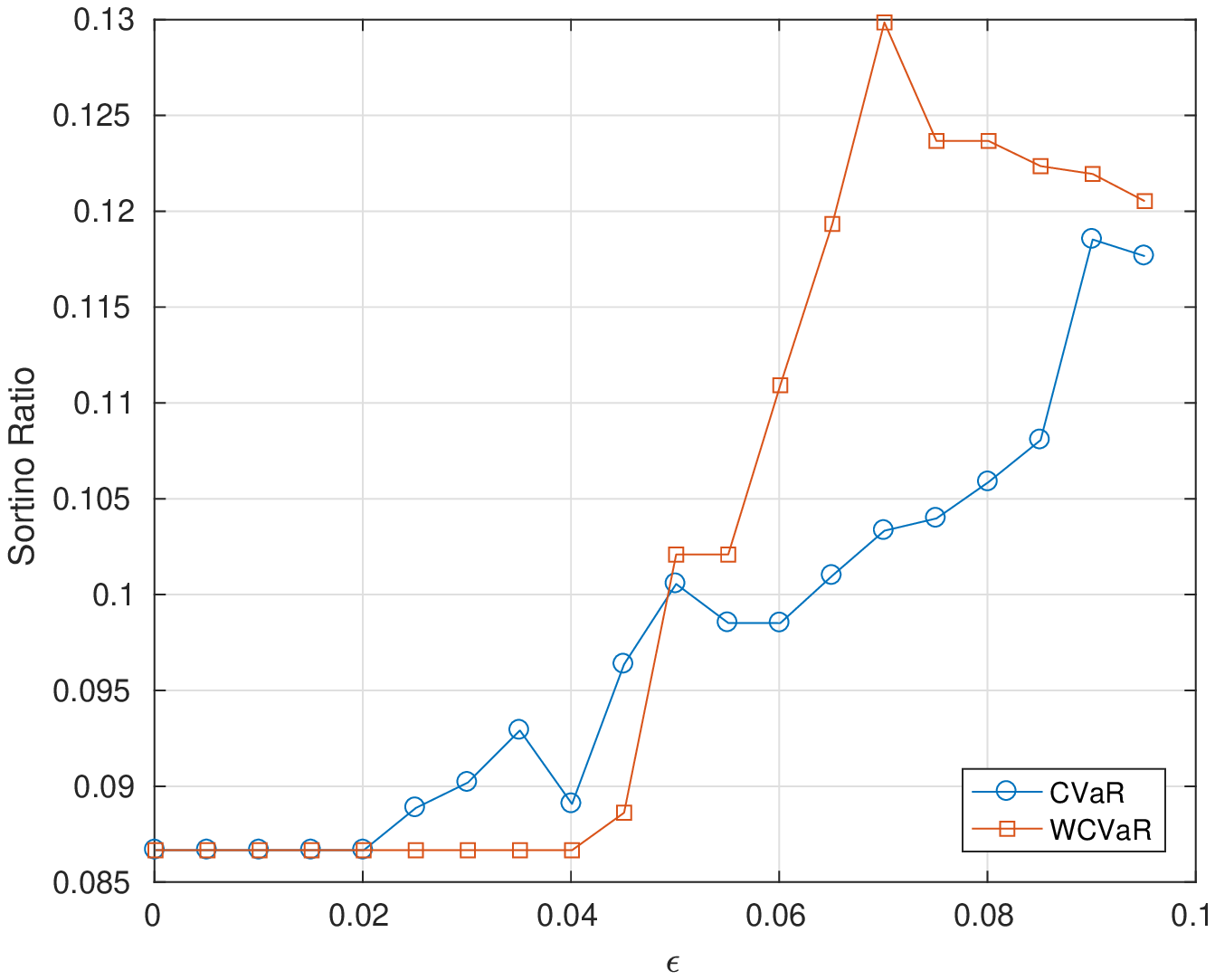}
\caption{Sortino ratio plot for base CVaR and WCVaR models in case of simulated data with $\zeta$ samples (31 assets) for $l=5$.}
\label{fig:6.2}
\end{figure}

On performing the simulation study with $\zeta$ samples, we draw an inference from Table \ref{avgtab:6.2} that the performance of the WCVaR model vis-a-vis the base CVaR model is maximal for $l=5$. Similar to the methodology followed in the previous case, we present the relevant results for $l=5$ in Figure \ref{fig:6.2} and Table \ref{tab:6.2}. It is difficult to draw any comparative inference from the Sortino Ratio plot of the two models in Figure \ref{fig:6.2}, since each outperforms the other in a different sub-interval of the range of $\epsilon$. The similar values of the Sortino Ratio in Table \ref{tab:6.2} as well as the marginal difference in the average Sortino Ratio for $l=5$ in Table \ref{avgtab:6.2} supports the claim of marginally superior performance of the WCVaR model with respect to the base CVaR model in this case.

\begin{figure}[!h]
\centering
\includegraphics[height=7.0cm]{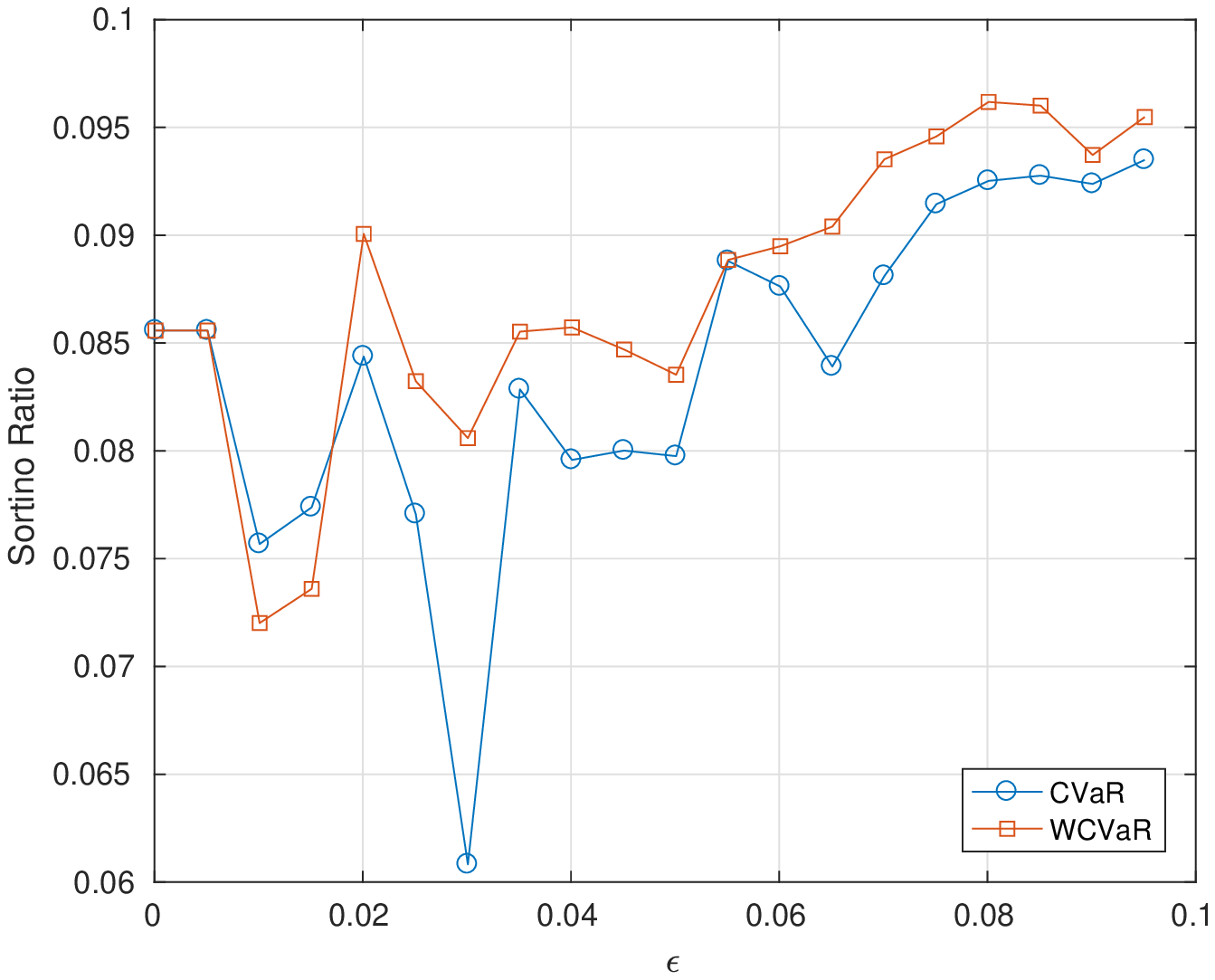}
\caption{Sortino ratio plot for base CVaR and WCVaR models in case of simulated data with $1000$ samples (31 assets) for $l=2$.}
\label{fig:6.3}
\end{figure}

The comparative analysis in terms of the average Sortino Ratio with the number of simulated samples being $1000$ is presented in Table \ref{avgtab:6.3}. Accordingly, we select the value of $l$ equal to 2 and perform empirical study of the base CVaR and WCVaR models, as presented in Figure \ref{fig:6.3} and Table \ref{tab:6.3}. From the plot of the Sortino Ratio in Figure \ref{fig:6.3}, we observe that the WCVaR model starts outperforming the base CVaR model after $\epsilon$ is close to $0.02$ \textit{i.e.,} incorporating robust optimization in CVaR minimization is advantageous in this case for investors who are not highly conservative. We infer that the performance of the WCVaR model is marginally better than that of the base CVaR model, which is evident from Table \ref{tab:6.3} as well.

\subsubsection{Performance with $N=98$ Assets}

\begin{figure}[!h]
\centering
\includegraphics[height=7.0cm]{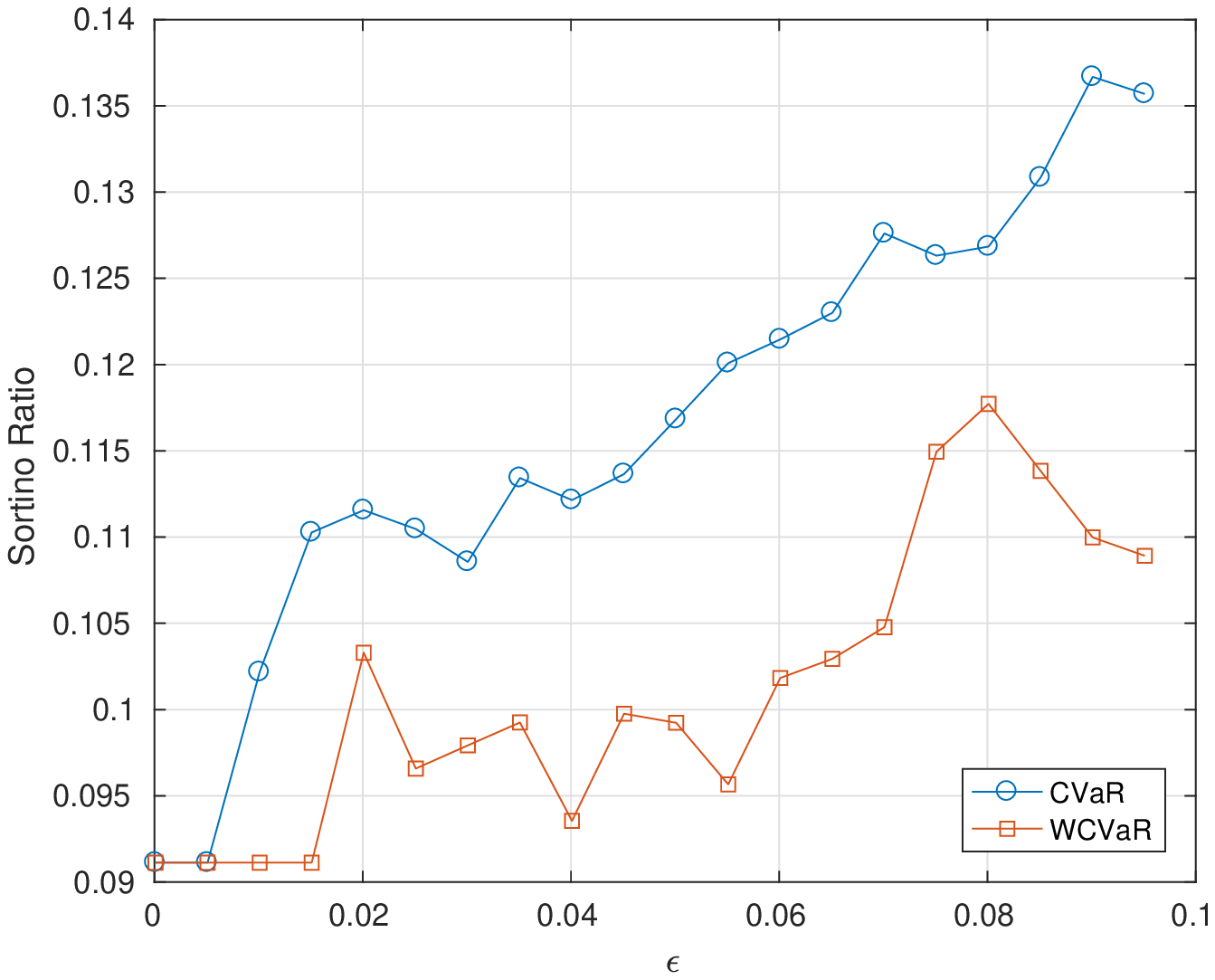}
\caption{Sortino ratio plot for base CVaR and WCVaR models in case of market data (98 assets) for $l=3$.}
\label{fig:6.4}
\end{figure}

In this subsection, we consider the scenario involving $N=98$ assets. Table \ref{avgtab:6.4} summarizes the comparative results observed using the base
CVaR and WCVaR models on the historical market data (involving stocks comprising S\&P BSE 100). Since the difference between the average Sortino Ratio of the
WCVaR and base CVaR models is maximum for $l=3$, so, we plot and tabulate the relevant results summarizing the empirical analysis of the two models for the same value of $l$ (Figure \ref{fig:6.4} and Table \ref{tab:6.4}). Similar to the corresponding case for the previous scenario, Figure \ref{fig:6.4} and Table \ref{tab:6.4} lead to an observation that the base CVaR model exhibits superior performance in comparison to the WCVaR model, taking into consideration the Sortino Ratio of the constructed portfolios as the performance measure.

\begin{figure}[!h]
\centering
\includegraphics[height=7.0cm]{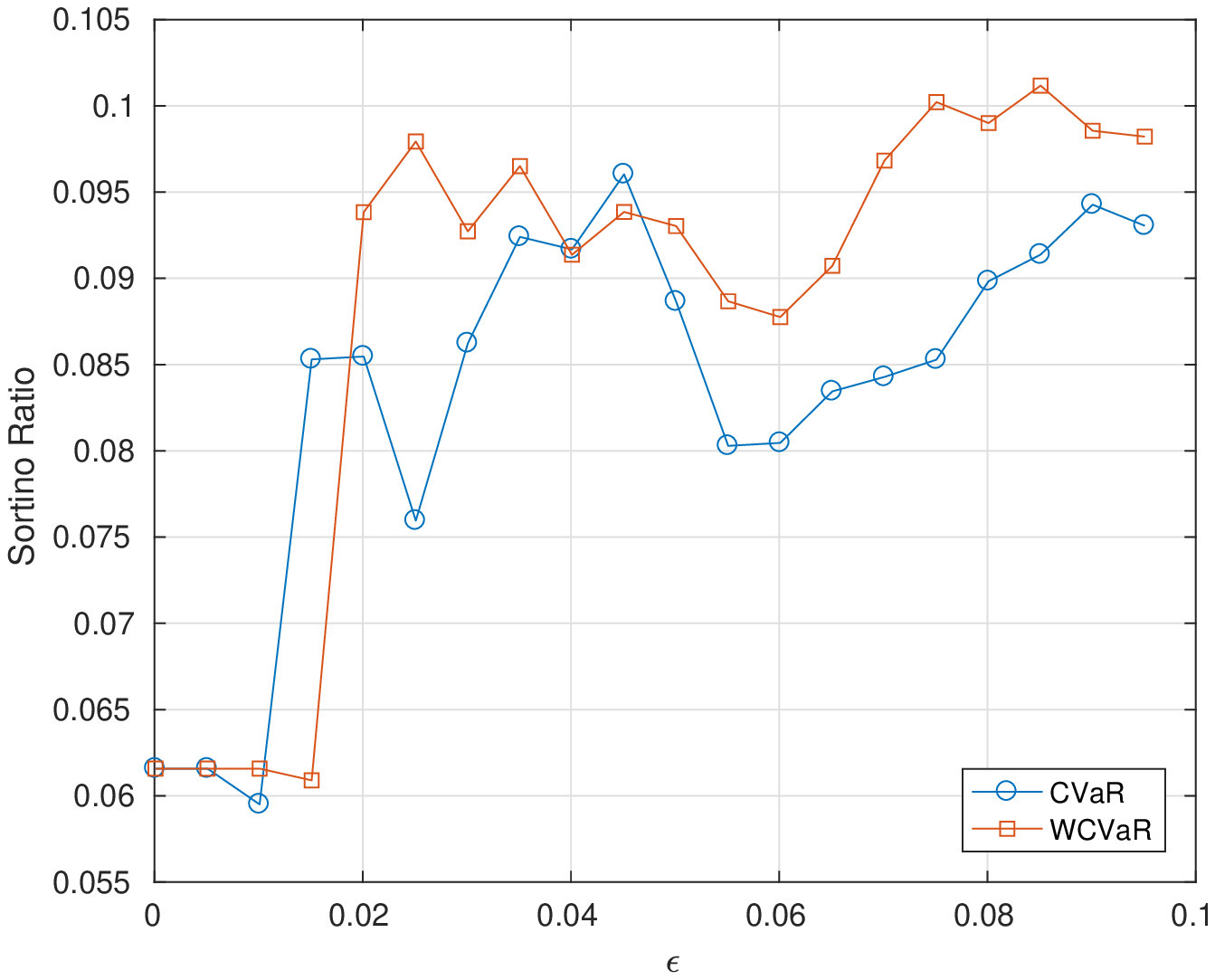}
\caption{Sortino ratio plot for base CVaR and WCVaR models in case of simulated data with $\zeta$ samples (98 assets) for $l=4$.}
\label{fig:6.5}
\end{figure}

Table \ref{avgtab:6.5} presents the comparison of the average Sortino Ratio for the simulation study with $\zeta$ samples \textit{i.e,} same number of samples as that of log-returns of S\&P BSE 100 data. In accordance with the discussed methodology, we choose the value of $l$ equal to $4$ and present the empirical results in Figure \ref{fig:6.5} and Table \ref{tab:6.5}. From Figure \ref{fig:6.5}, we observe that the WCVaR model outperforms the base CVaR model in terms of the Sortino Ratio for majority of the values of $\epsilon \in (0,0.1)$. From Table \ref{tab:6.5}, we can support this inference quantitatively by observing the greater Sortino Ratio of the portfolios for the WCVaR model vis-a-vis the base CVaR model in most of the cases.

\begin{figure}[!h]
\centering
\includegraphics[height=7.0cm]{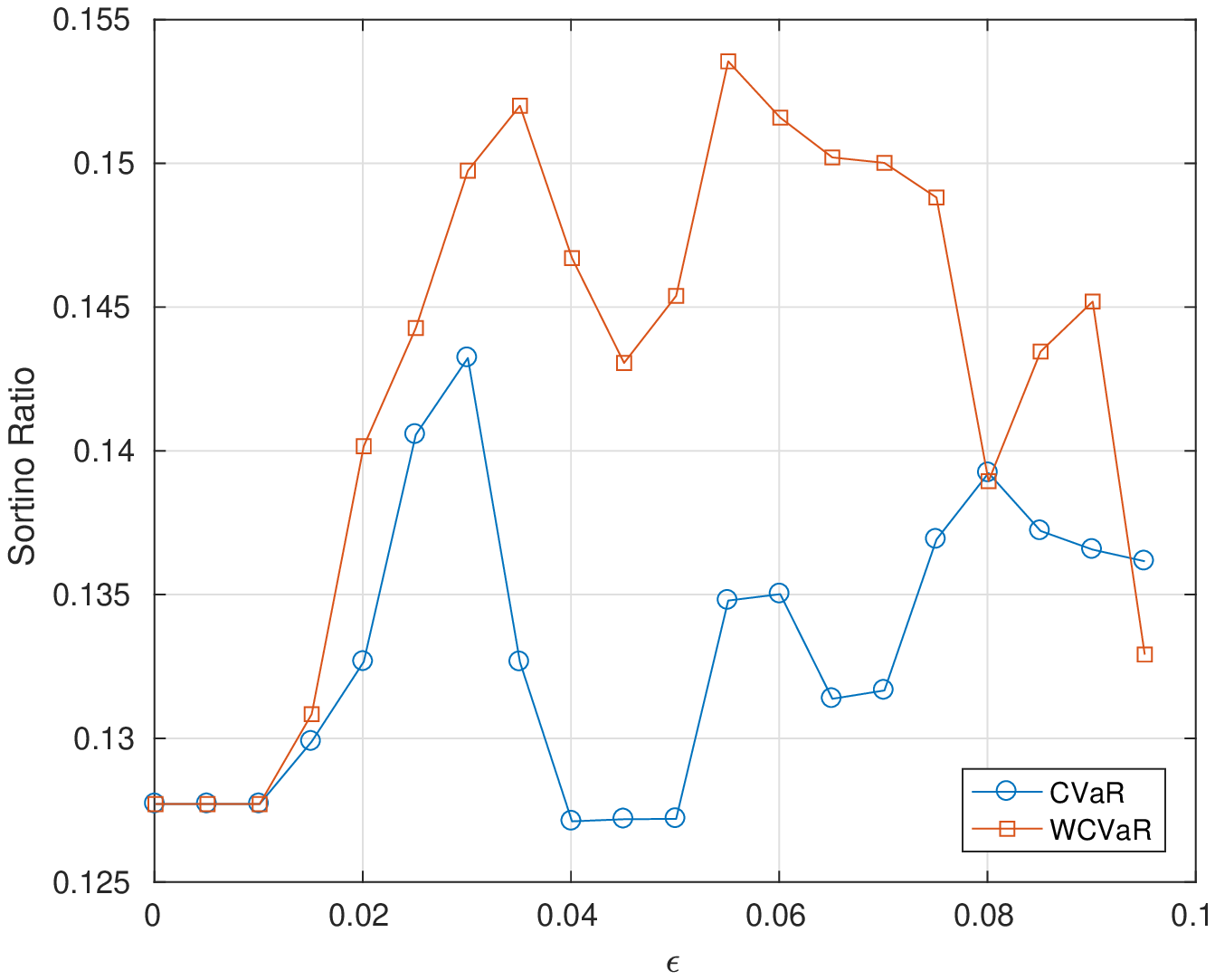}
\caption{Sortino ratio plot for base CVaR and WCVaR models in case of simulated data with $1000$ samples (98 assets) for $l=5$.}
\label{fig:6.6}
\end{figure}

Finally, the comparative analysis on the basis of the average Sortino Ratio for the simulated data with $1000$ samples is presented in Table \ref{avgtab:6.6}. Accordingly, the results corresponding to the empirical study for $l=5$ are presented in Figure \ref{fig:6.6} and Table \ref{tab:6.6}. As observed from the plot of the Sortino Ratio in Figure \ref{fig:6.6},
the WCVaR model performs superior with respect to the base CVaR model almost entirely for $\epsilon \in (0,0.1)$. This observation is supported by the results presented in Table \ref{tab:6.6}.

\section{Discussion}
\label{Discussion}

We now present the discussion regarding the performance of the WVaR and the WCVaR models, with respect to their classical counterparts using both qualitative and qualitative arguments.

\subsection{Robust Optimization in VaR Minimization}

We discuss regarding the practical viability of incorporating robust optimization in VaR minimization from the point of view of number of stocks, sample size and types of data. For the sake of convenience, we tabulate all the results obtained in preceding Sections in Table \ref{tab:var_conc}, where for a particular scenario, we tabulated the average Sortino Ratio obtained over the chosen range of $\epsilon$.

\subsubsection{From the Standpoint of Number of Stocks}

From Table \ref{tab:var_conc}, we observe a common inference that the WVaR model exhibits superior (inferior) performance than the base VaR model in the case of $N=98$ ($N=31$). The qualitative argument for this kind of behaviour cannot be attributed to the diversification of the portfolio because at times, VaR may not be sub-additive. We justify this behaviour on the lines of Michaud \cite{Michaud} and Ghaoui et al. \cite{Ghaoui03}. As Michaud points out, the errors in the estimation of mean and covariances of the asset returns accumulates as the number of stocks increase. This implies that the data uncertainty in the case of larger number of stocks is more in comparison to smaller number of stocks. The WVaR model can handle the data uncertainty in a better way than the base VaR model \cite{Ghaoui03}. Therefore, we observe an increment in the Sortino Ratio for both the models, but the increment is more in case of WVaR model such that it outperforms the base VaR model in all types of data environments. In the scenarios involving the sets of simulated data, the same argument explains the behaviour, as the estimated moments' pair are used as true moments' pair, for the generation of the data. So the error in the estimation of the mean and the variance of the asset returns affects the behaviour in the same way as in the market data. However, in the case of simulated environment of smaller number of stocks ($N=31$) with $1000$ simulated samples, the WVaR model outperforms the base VaR model. We qualitatively justify this kind of exceptional behaviour in the next section.

\subsubsection{From the Standpoint of Number of Simulations}

When we observe the results from the perspective of number of simulations, we can draw some insightful conclusions. For the case $N=98$, the better performance of WVaR model can be attributed to the reason above. But, when $N=31$, one can observe that the performance of the optimal portfolio, when $1000$ samples were simulated is better than that of the portfolio obtained when $\zeta$ number of samples were simulated. The reason for this sort of behaviour lies in the subroutine of the Non Parametric Bootstrap Algorithm, that uses sampling with ``replacement''. Therefore, more the number of samples, better are the bounds that one can obtain from the algorithm. So, when we generate more number of samples ($1000$), the WVaR model which uses these bounds, performs better than the case where $\zeta$ number of samples are used for computing the bounds. In this setup, the increment in the number of simulation transits the robust portfolio from under-performing ($\zeta$ case) to out-performing ($1000$ case) the portfolio obtained from base VaR model.

\subsubsection{From the Standpoint of Type of the Data}

As explained in the previous section, when $N=31$, the superior or equivalent performance of the WVaR model over the base VaR model in case of simulated data can be attributed to the following reason: The real market data is difficult to model and may not follow any distribution, whereas the simulated data follows the multivariate normal distribution with their true moments' pair as the tuple of estimated mean and covariance matrix of the asset returns from the real market data. Therefore, both the base VaR and the WVaR models exhibits superior performance in the case of real market data than the simulated data, because the real market data has more data uncertainty. The reason for the out-performance of WVaR model over the base VaR model for $N=98$ stocks has already been discussed in the preceding sections.

\subsection{Robust Optimization in CVaR Minimization}

Similar to VaR minimization, we conduct the performance analysis of the WCVaR model with respect to its base case counterpart from different standpoints. Table \ref{tab:cvar_conc} presents the average Sortino Ratio of the base CVaR and the WCVaR models for each scenario based on the chosen value of $l$, as per the methodology discussed in the preceding Section.

\subsubsection{From the Standpoint of Number of Stocks}

We begin with a discussion of the results presented in Table \ref{tab:cvar_conc} from the standpoint of number of stocks. We observe a favorable trend for the case of simulated data with $1000$ samples. For $N=31$, the WCVaR model performs marginally better than the base CVaR model. As $N$ increases to $98$, the WCVaR model exhibits superior performance vis-a-vis the base CVaR model. Since CVaR and WCVaR are coherent risk measures, so, increase in the number of stocks enhances diversification. As a result, an uptrend is observed in the performance of these two models for $N=98$. But, WCVaR, being a robust risk measure, diversifies over worst-case scenarios as well through mixture distribution uncertainty, as defined in equation (\ref{eq:6.8}). As a result, the performance of the WCVaR model, with respect to the base CVaR model enhances when larger number of stocks are taken into consideration.

For the case involving market data, we observe that the base CVaR model performs better than the WCVaR model in the scenario involving less number of stocks ($N=31$). Even after increasing $N$ to 98, we draw the same inference. The reason behind such kind of opposite trend can be attributed to the lack of knowledge regarding the distribution of the returns in the real market data. Computation of CVaR assumes that the probability distribution is perfectly known and optimization of WCVaR is based on the assumption of the return distribution belonging to a mixture of some prior likelihood distributions. Since the market data hardly follows any distribution, so, there is a sense of ambiguity associated with optimizing CVaR and WCVaR using the discrete sampling technique (discussed in Section \ref{Risk_Minimization_Approaches}) for the market data.

However, in the case of simulated data with $\zeta$ samples, an unusual trend is observed, despite involving comparative inference similar to the previous case. On increasing $N$ to 98, we observe a decline in the performance of the base CVaR and WCVaR models. The reason behind such observation is not obvious.

\subsubsection{From the Standpoint of Number of Simulations}

We now compare the performance of the two models based on the number of simulated samples. From Table \ref{tab:cvar_conc}, for the case involving $\zeta$ simulations, the WCVaR model exhibits superior or equivalent performance with respect to the base CVaR model irrespective of the number of stocks. Same inference can be drawn with $1000$ simulated samples as well.

\subsubsection{From the Standpoint of Type of the Data}

Due to similar reasons as in VaR minimization, an opposite trend is observed in the case of real market data (as discussed above) when the number of stocks is less ($N=31$). Similar observation is inferred for the market data on taking into account the larger number of stocks ($N=98$).

\section{Conclusion}
\label{Conclusion}

Akin to mean variance analysis, there is a problem of lack of robustness in the classical formulations of VaR and CVaR minimization. We discuss and assess the performance of the robust counterparts for these optimization problems that have been formulated to address this concern. Motivated by the results by Ghaoui et al. \cite{Ghaoui03}, we formulate the worst case robust version of the VaR model using separable uncertainty set. Regardless of the type of the data, be it from real market or from a simulated environment, we observe favourable results for the WVaR model with Sortino ratio as the performance measure when the portfolio comprises higher number of stocks. Such an outperformance can also be achieved in case of simulated data with larger sample size.

In contrast to the results reported by Zhu \cite{Zhu}, we observe that the base CVaR model performs better than the robust counterpart (formulated by incorporating mixture distribution uncertainty) in the case of Market data irrespective of the number of stocks comprising in the optimal portfolio. This could be attributed to the following two reasons:
\begin{itemize}
\item Incorporation of different weight constraints in our optimization problem.
\item In contrast to \cite{Zhu}, our work uses Sortino Ratio as a performance measure.
\end{itemize}
On the other hand, in the case of simulated data, we draw a favourable inference by noting superior or equivalent performance of the WCVaR model vis-a-vis the base CVaR model. In accordance with these results, we advocate for consideration of worst case models as a viable alternative to their classical counterparts especially in the case of higher number of stocks and in a simulated environment.

\bibliographystyle{abbrv}
\bibliography{bib}

\newpage

\begin{table}[!h]
\centering
\captionsetup{justification=centering}
\begin{tabular}{||c|c|c|c|c|c|c||}
\hline
$\epsilon$ & $\mu_{VaR}$ & $\sigma_{VaR}^{d}$ & $\mu_{WVaR}$ & $\sigma_{WVaR}^{d}$ & $SR_{VaR}$ & $SR_{WVaR}$\\
\hline
0.0001 & 0.000646 & 0.00571 & 0.0006 & 0.00557 & 0.0851 & 0.0791 \\
0.0201 & 0.000715 & 0.00572 & 0.000657 & 0.00556 & 0.0971 & 0.0895 \\
0.0401 & 0.000744 & 0.00572 & 0.000684 & 0.00556 & 0.102 & 0.0942 \\
0.0601 & 0.000763 & 0.00573 & 0.000705 & 0.00556 & 0.105 & 0.098 \\
0.0801 & 0.000777 & 0.00573 & 0.000723 & 0.00556 & 0.108 & 0.101 \\
\hline
& & & & Avg & 0.102 & 0.0946 \\
\hline
\end{tabular}
\caption{Empirical analysis of base VaR and WVaR models in case of market data (31 assets).}
\label{tab:5.1}
\end{table}

\begin{table}[!h]
\centering
\captionsetup{justification=centering}
\begin{tabular}{||c|c|c|c|c|c|c||}
\hline
$\epsilon$ & $\mu_{VaR}$ & $\sigma_{VaR}^{d}$ & $\mu_{WVaR}$ & $\sigma_{WVaR}^{d}$ & $SR_{VaR}$ & $SR_{WVaR}$\\
\hline
0.0001 & 0.000444 & 0.00571 & 0.000417 & 0.00548 & 0.0497 & 0.0469 \\
0.0201 & 0.000562 & 0.0057 & 0.000506 & 0.00548 & 0.0706 & 0.0632 \\
0.0401 & 0.00061 & 0.0057 & 0.000554 & 0.00548 & 0.079 & 0.0719 \\
0.0601 & 0.000647 & 0.00571 & 0.000592 & 0.00548 & 0.0855 & 0.0788 \\
0.0801 & 0.000681 & 0.00571 & 0.000625 & 0.00549 & 0.0911 & 0.0849 \\
\hline
& & & & Avg & 0.0793 & 0.0728 \\
\hline
\end{tabular}
\caption{Empirical analysis of base VaR and WVaR models in case of simulated data with $\zeta$ number of samples (31 assets).}
\label{tab:5.2}
\end{table}

\begin{table}[!h]
\centering
\captionsetup{justification=centering}
\begin{tabular}{||c|c|c|c|c|c|c||}
\hline
$\epsilon$ & $\mu_{VaR}$ & $\sigma_{VaR}^{d}$ & $\mu_{WVaR}$ & $\sigma_{WVaR}^{d}$ & $SR_{VaR}$ & $SR_{WVaR}$\\
\hline
0.0001 & 0.000587 & 0.00578 & 0.000605 & 0.00568 & 0.0739 & 0.0784 \\
0.0201 & 0.000636 & 0.00577 & 0.000645 & 0.00568 & 0.0824 & 0.0855 \\
0.0401 & 0.00066 & 0.00577 & 0.000663 & 0.00567 & 0.0866 & 0.0888 \\
0.0601 & 0.00068 & 0.00577 & 0.000678 & 0.00567 & 0.0901 & 0.0914 \\
0.0801 & 0.000694 & 0.00578 & 0.000691 & 0.00567 & 0.0925 & 0.0936 \\
\hline
& & & & Avg & 0.0869 & 0.089 \\
\hline
\end{tabular}
\caption{Empirical analysis of base VaR and WVaR models in case of simulated data with $1000$ samples (31 assets).}
\label{tab:5.3}
\end{table}

\begin{table}[!h]
\centering
\captionsetup{justification=centering}
\begin{tabular}{||c|c|c|c|c|c|c||}
\hline
$\epsilon$ & $\mu_{VaR}$ & $\sigma_{VaR}^{d}$ & $\mu_{WVaR}$ & $\sigma_{WVaR}^{d}$ & $SR_{VaR}$ & $SR_{WVaR}$\\
\hline
0.0001 & 0.000667 & 0.00553 & 0.000711 & 0.00531 & 0.0918 & 0.104 \\
0.0201 & 0.000713 & 0.00552 & 0.000757 & 0.00529 & 0.1 & 0.113 \\
0.0401 & 0.000735 & 0.00551 & 0.000778 & 0.00529 & 0.104 & 0.117 \\
0.0601 & 0.000751 & 0.00551 & 0.000795 & 0.00528 & 0.107 & 0.12 \\
0.0801 & 0.000765 & 0.0055 & 0.000809 & 0.00527 & 0.11 & 0.123 \\
\hline
& & & & Avg & 0.105 & 0.117 \\
\hline
\end{tabular}
\caption{Empirical analysis of base VaR and WVaR models in case of market data (98 assets).}
\label{tab:5.4}
\end{table}

\begin{table}[!h]
\centering
\captionsetup{justification=centering}
\begin{tabular}{||c|c|c|c|c|c|c||}
\hline
$\epsilon$ & $\mu_{VaR}$ & $\sigma_{VaR}^{d}$ & $\mu_{WVaR}$ & $\sigma_{WVaR}^{d}$ & $SR_{VaR}$ & $SR_{WVaR}$\\
\hline
0.0001 & 0.000341 & 0.00599 & 0.00052 & 0.00549 & 0.0302 & 0.0655 \\
0.0201 & 0.000434 & 0.00594 & 0.000629 & 0.00546 & 0.0461 & 0.086 \\
0.0401 & 0.000475 & 0.00593 & 0.00067 & 0.00545 & 0.0532 & 0.0936 \\
0.0601 & 0.000508 & 0.00592 & 0.000697 & 0.00544 & 0.0589 & 0.0988 \\
0.0801 & 0.000537 & 0.00591 & 0.000719 & 0.00544 & 0.0639 & 0.103 \\
\hline
& & & & Avg & 0.0538 & 0.0932 \\
\hline
\end{tabular}
\caption{Empirical analysis of base VaR and WVaR models in case of simulated data with $\zeta$ number of samples (98 assets).}
\label{tab:5.5}
\end{table}

\begin{table}[!h]
\centering
\captionsetup{justification=centering}
\begin{tabular}{||c|c|c|c|c|c|c||}
\hline
$\epsilon$ & $\mu_{VaR}$ & $\sigma_{VaR}^{d}$ & $\mu_{WVaR}$ & $\sigma_{WVaR}^{d}$ & $SR_{VaR}$ & $SR_{WVaR}$\\
\hline
0.0001 & 0.000702 & 0.00596 & 0.000822 & 0.00559 & 0.091 & 0.119 \\
0.0201 &0.000772 & 0.00592 & 0.000897 & 0.00555 & 0.103 & 0.133 \\
0.0401 & 0.000803 & 0.00591 & 0.000931 & 0.00555 & 0.109 & 0.139 \\
0.0601 & 0.000828 & 0.0059 & 0.000959 & 0.00554 & 0.113 & 0.144 \\
0.0801 & 0.00085 & 0.00589 & 0.000982 & 0.00553 & 0.117 & 0.149 \\
\hline
& & & & Avg & 0.109 & 0.14 \\
\hline
\end{tabular}
\caption{Empirical analysis of base VaR and WVaR models in case of simulated data with $1000$ samples (98 assets).}
\label{tab:5.6}
\end{table}

\clearpage

\begin{table}[!h]
\centering
\captionsetup{justification=centering}
\begin{tabular}{||c|c|c|c||}
\hline
$l$ & $Avg. \, \, SR_{CVaR}$ & $Avg. \, \, SR_{WCVaR}$ & $Diff. \, \, in \, \, Avg. \, \, SR$ \\
\hline
2 & 0.084 & 0.0626 & -0.0214 \\
3 & 0.084 & 0.036 & -0.048 \\
4 & 0.084 & 0.0455 & -0.0385 \\
5 & 0.084 & 0.0327 & -0.0513 \\
\hline
\end{tabular}
\caption{Comparison of base CVaR and WCVaR in case of market data (31 assets) for different values of $l$.}
\label{avgtab:6.1}
\end{table}

\begin{table}[!h]
\centering
\captionsetup{justification=centering}
\begin{tabular}{||c|c|c|c|c|c|c||}
\hline
$\epsilon$ & $\mu_{CVaR}$ & $\sigma_{CVaR}^{d}$ & $\mu_{WCVaR}$ & $\sigma_{WCVaR}^{d}$ & $SR_{CVaR}$ & $SR_{WCVaR}$\\
\hline
0.0001 & 0.000266 & 0.00599 & 0.000266 & 0.00599 & 0.0178 & 0.0178 \\
0.0201 & 0.000545 & 0.00604 & 0.000266 & 0.00599 & 0.0639 & 0.0178 \\
0.0401 & 0.000706 & 0.00596 & 0.000514 & 0.00586 & 0.0918 & 0.0604 \\
0.0601 & 0.000832 & 0.00587 & 0.000645 & 0.00587 & 0.115 & 0.0828 \\
0.0801 & 0.000886 & 0.00582 & 0.000696 & 0.00587 & 0.125 & 0.0914 \\
\hline
\end{tabular}
\caption{Empirical analysis of base CVaR and WCVaR models in case of market data (31 assets) for $l=2$.}
\label{tab:6.1}
\end{table}

\begin{table}[!h]
\centering
\captionsetup{justification=centering}
\begin{tabular}{||c|c|c|c||}
\hline
$l$ & $Avg. \, \, SR_{CVaR}$ & $Avg. \, \, SR_{WCVaR}$ & $Diff. \, \, in \, \, Avg. \, \, SR$ \\
\hline
2 & 0.0973 & 0.0973 & -5.78e-05 \\
3 & 0.0973 & 0.0974 & 6.59e-05 \\
4 & 0.0973 & 0.0998 & 0.0025 \\
5 & 0.0973 & 0.102 & 0.00492 \\
\hline
\end{tabular}
\caption{Comparison of base CVaR and WCVaR models in case of simulated data with $\zeta$ samples (31 assets) for different values of $l$.}
\label{avgtab:6.2}
\end{table}

\begin{table}[!h]
\centering
\captionsetup{justification=centering}
\begin{tabular}{||c|c|c|c|c|c|c||}
\hline
$\epsilon$ & $\mu_{CVaR}$ & $\sigma_{CVaR}^{d}$ & $\mu_{WCVaR}$ & $\sigma_{WCVaR}^{d}$ & $SR_{CVaR}$ & $SR_{WCVaR}$\\
\hline
0.0001 & 0.000675 & 0.00595 & 0.000675 & 0.00595 & 0.0867 & 0.0867 \\
0.0201 & 0.000675 & 0.00595 & 0.000675 & 0.00595 & 0.0867 & 0.0867 \\
0.0401 & 0.000694 & 0.006 & 0.000675 & 0.00595 & 0.0891 & 0.0867 \\
0.0601 & 0.000731 & 0.0058 & 0.000838 & 0.00611 & 0.0985 & 0.111 \\
0.0801 & 0.000776 & 0.00582 & 0.000889 & 0.0059 & 0.106 & 0.124 \\
\hline
\end{tabular}
\caption{Empirical analysis of base CVaR and WCVaR models in case of simulated data with $\zeta$ samples (31 assets) for $l=5$.}
\label{tab:6.2}
\end{table}

\begin{table}[!h]
\centering
\captionsetup{justification=centering}
\begin{tabular}{||c|c|c|c||}
\hline
$l$ & $Avg. \, \, SR_{CVaR}$ & $Avg. \, \, SR_{WCVaR}$ & $Diff. \, \, in \, \, Avg. \, \, SR$ \\
\hline
2 & 0.084 & 0.0874 & 0.00343 \\
3 & 0.084 & 0.0855 & 0.00147 \\
4 & 0.084 & 0.0874 & 0.00337 \\
5 & 0.084 & 0.0864 & 0.00238 \\
\hline
\end{tabular}
\caption{Comparison of base CVaR and WCVaR models in case of simulated data with $1000$ samples (31 assets) for different values of $l$.}
\label{avgtab:6.3}
\end{table}

\clearpage

\begin{table}[!h]
\centering
\captionsetup{justification=centering}
\begin{tabular}{||c|c|c|c|c|c|c||}
\hline
$\epsilon$ & $\mu_{CVaR}$ & $\sigma_{CVaR}^{d}$ & $\mu_{WCVaR}$ & $\sigma_{WCVaR}^{d}$ & $SR_{CVaR}$ & $SR_{WCVaR}$\\
\hline
0.0001 & 0.000677 & 0.00604 & 0.000677 & 0.00604 & 0.0856 & 0.0856 \\
0.0201 & 0.000658 & 0.0059 & 0.000691 & 0.0059 & 0.0844 & 0.0901 \\
0.0401 & 0.000629 & 0.0059 & 0.000663 & 0.00587 & 0.0796 & 0.0857 \\
0.0601 & 0.00067 & 0.00582 & 0.000682 & 0.00584 & 0.0876 & 0.0895 \\
0.0801 & 0.000701 & 0.00585 & 0.00072 & 0.00583 & 0.0925 & 0.0962 \\
\hline
\end{tabular}
\caption{Empirical analysis of base CVaR and WCVaR models in case of simulated data with $1000$ samples (31 assets) for $l=2$.}
\label{tab:6.3}
\end{table}

\begin{table}[!h]
\centering
\captionsetup{justification=centering}
\begin{tabular}{||c|c|c|c||}
\hline
$l$ & $Avg. \, \, SR_{CVaR}$ & $Avg. \, \, SR_{WCVaR}$ & $Diff. \, \, in \, \, Avg. \, \, SR$ \\
\hline
2 & 0.116 & 0.0984 & -0.0181 \\
3 & 0.116 & 0.101 & -0.0153 \\
4 & 0.116 & 0.0967 & -0.0198 \\
5 & 0.116 & 0.0912 & -0.0253 \\
\hline
\end{tabular}
\caption{Comparison of base CVaR and WCVaR models in case of market data (98 assets) for different values of $l$.}
\label{avgtab:6.4}
\end{table}

\begin{table}[!h]
\centering
\captionsetup{justification=centering}
\begin{tabular}{||c|c|c|c|c|c|c||}
\hline
$\epsilon$ & $\mu_{CVaR}$ & $\sigma_{CVaR}^{d}$ & $\mu_{WCVaR}$ & $\sigma_{WCVaR}^{d}$ & $SR_{CVaR}$ & $SR_{WCVaR}$\\
\hline
0.0001 & 0.000687 & 0.00579 & 0.000687 & 0.00579 & 0.0911 & 0.0911 \\
0.0201 & 0.000786 & 0.00561 & 0.000755 & 0.00576 & 0.112 & 0.103 \\
0.0401 & 0.00079 & 0.00562 & 0.000692 & 0.00569 & 0.112 & 0.0935 \\
0.0601 & 0.000839 & 0.0056 & 0.000738 & 0.00567 & 0.121 & 0.102 \\
0.0801 & 0.000847 & 0.00542 & 0.00082 & 0.00561 & 0.127 & 0.118 \\
\hline
\end{tabular}
\caption{Empirical analysis of base CVaR and WCVaR models in case of market data (98 assets) for $l=3$.}
\label{tab:6.4}
\end{table}

\begin{table}[!h]
\centering
\captionsetup{justification=centering}
\begin{tabular}{||c|c|c|c||}
\hline
$l$ & $Avg. \, \, SR_{CVaR}$ & $Avg. \, \, SR_{WCVaR}$ & $Diff. \, \, in \, \, Avg. \, \, SR$ \\
\hline
2 & 0.0833 & 0.0854 & 0.00209 \\
3 & 0.0833 & 0.0844 & 0.00102 \\
4 & 0.0833 & 0.0883 & 0.00497 \\
5 & 0.0833 & 0.0779 & -0.00543 \\
\hline
\end{tabular}
\caption{Comparison of base CVaR and WCVaR models in case of simulated data with $\zeta$ samples (98 assets) for different values of $l$.}
\label{avgtab:6.5}
\end{table}

\begin{table}[!h]
\centering
\captionsetup{justification=centering}
\begin{tabular}{||c|c|c|c|c|c|c||}
\hline
$\epsilon$ & $\mu_{CVaR}$ & $\sigma_{CVaR}^{d}$ & $\mu_{WCVaR}$ & $\sigma_{WCVaR}^{d}$ & $SR_{CVaR}$ & $SR_{WCVaR}$\\
\hline
0.0001 & 0.000554 & 0.0064 & 0.000554 & 0.0064 & 0.0616 & 0.0616 \\
0.0201 & 0.000703 & 0.00636 & 0.000755 & 0.00635 & 0.0855 & 0.0938 \\
0.0401 & 0.000727 & 0.00619 & 0.000723 & 0.00617 & 0.0917 & 0.0914 \\
0.0601 & 0.000648 & 0.00607 & 0.000698 & 0.00613 & 0.0805 & 0.0878 \\
0.0801 & 0.000706 & 0.00609 & 0.000768 & 0.00614 & 0.0898 & 0.099 \\
\hline
\end{tabular}
\caption{Empirical analysis of base CVaR and WCVaR models in case of simulated data with $\zeta$ samples (98 assets) for $l=4$.}
\label{tab:6.5}
\end{table}

\begin{table}[!h]
\centering
\captionsetup{justification=centering}
\begin{tabular}{||c|c|c|c||}
\hline
$l$ & $Avg. \, \, SR_{CVaR}$ & $Avg. \, \, SR_{WCVaR}$ & $Diff. \, \, in \, \, Avg. \, \, SR$ \\
\hline
2 & 0.133 & 0.133 & -0.000273 \\
3 & 0.133 & 0.137 & 0.00422 \\
4 & 0.133 & 0.139 & 0.00624 \\
5 & 0.133 & 0.142 & 0.00937 \\
\hline
\end{tabular}
\caption{Comparison of base CVaR and WCVaR models in case of simulated data with $1000$ samples (98 assets) for different values of $l$.}
\label{avgtab:6.6}
\end{table}

\begin{table}[!h]
\centering
\captionsetup{justification=centering}
\begin{tabular}{||c|c|c|c|c|c|c||}
\hline
$\epsilon$ & $\mu_{CVaR}$ & $\sigma_{CVaR}^{d}$ & $\mu_{WCVaR}$ & $\sigma_{WCVaR}^{d}$ & $SR_{CVaR}$ & $SR_{WCVaR}$\\
\hline
0.0001 & 0.000935 & 0.00607 & 0.000935 & 0.00607 & 0.128 & 0.128 \\
0.0201 & 0.00097 & 0.00611 & 0.00101 & 0.00609 & 0.133 & 0.14 \\
0.0401 & 0.000927 & 0.00604 & 0.00104 & 0.00603 & 0.127 & 0.147 \\
0.0601 & 0.000963 & 0.00595 & 0.00106 & 0.00595 & 0.135 & 0.152 \\
0.0801 & 0.000974 & 0.00585 & 0.000981 & 0.00591 & 0.139 & 0.139 \\
\hline
\end{tabular}
\caption{Empirical analysis of base CVaR and WCVaR models in case of simulated data with $1000$ samples (98 assets) for $l=5$.}
\label{tab:6.6}
\end{table}

\begin{table}[!h]
\centering
\small
\captionsetup{justification=centering}
\begin{tabular}{|c|c|c|c|c|c|c|}
\hline
\multirow{2}{*}{} $N$ &
\multicolumn{3}{c|}{$N=31$} &
\multicolumn{3}{c|}{$N=98$}  \\
\hline
Type of & Market & Sim. data & Sim. data & Market & Sim. data & Sim. data \\
data & data & $\zeta$ samples & $1000$ samples & data & $\zeta$ samples & $1000$ samples \\
\hline
VaR & 0.102 & 0.0793 & 0.0869 & 0.105 & 0.0538 & 0.109 \\
\hline
WVaR & 0.0946 & 0.0728 & 0.089 & 0.117 & 0.0932 & 0.14 \\
\hline
\end{tabular}
\caption{Comparison of the average Sortino ratio for the base VaR and WVaR models in various scenarios.}
\label{tab:var_conc}
\end{table}

\begin{table}[!h]
\centering
\small
\captionsetup{justification=centering}
\begin{tabular}{|c|c|c|c|c|c|c|}
\hline
\multirow{2}{*}{} $N$ &
\multicolumn{3}{c|}{$N=31$} &
\multicolumn{3}{c|}{$N=98$}  \\
\hline
Type of & Market & Sim. data & Sim. data & Market & Sim. data & Sim. data \\
data & data & $\zeta$ samples & $1000$ samples & data & $\zeta$ samples & $1000$ samples \\
\hline
CVaR & 0.084 & 0.0973 & 0.084 & 0.116 & 0.0833 & 0.133 \\
\hline
WCVaR & 0.0626 & 0.102 & 0.0874 & 0.101 & 0.0883 & 0.142 \\
\hline
\end{tabular}
\caption{Comparison of the average Sortino ratio for the base CVaR and WCVaR models in various scenarios.}
\label{tab:cvar_conc}
\end{table}

\end{document}